\documentclass[fleqn,usenatbib]{mnras}
\usepackage{url}
\usepackage{newtxtext,newtxmath}
\usepackage{comment}
\usepackage{pdflscape}

\usepackage[T1]{fontenc}
 \usepackage{amsmath}
\DeclareRobustCommand{\VAN}[3]{#2}
\let\VANthebibliography\thebibliography
\def\thebibliography{\DeclareRobustCommand{\VAN}[3]{##3}\VANthebibliography}

\usepackage{graphicx}	
\usepackage{amsmath}	

\title[The nature of extreme red-\textit{Herschel} galaxies]{ALMA follow-up of $\sim$ 3,000 red-Herschel  galaxies: the nature of extreme submillimeter galaxies.}

\author[
Marianela Quirós-Rojas et al.]{
Marianela Quirós-Rojas,$^{1}$\thanks{E-mail: mquiros@inaoe.mx}
Alfredo Montaña,$^{1}$
Jorge A. Zavala,$^{2}$
Itziar Aretxaga,$^{1}$
and David H. Hughes$^{1}$.
\\
$^{1}$Instituto Nacional de Astrof\'{i}sica \'{O}ptica y Electr\'{o}nica, Luis Enrique Erro 1, Tonantzintla CP 72840, Puebla, M\'{e}xico\\
$^{2}$National Astronomical Observatory of Japan, 2-21-1 Osawa, Mitaka, Tokyo 181-8588, Japan\\
}

\date{Accepted XXX. Received YYY; in original form ZZZ}

\pubyear{2024}

\begin{document}
\label{firstpage}
\pagerange{\pageref{firstpage}--\pageref{lastpage}}
\maketitle

\begin{abstract}
We present the analysis of over 3,000 red-\textit{Herschel} sources ($S_{\mathrm{250\mu m}}<S_{\mathrm{350\mu m}}<S_{\mathrm{500\mu m} }$) using public data from the ALMA archive and the \textit{Herschel}-ATLAS survey. This represents the largest sample of red-\textit{Herschel} sources with interferometric follow-up observations to date. The high ALMA angular resolution and sensitivity ($\theta_{\rm FWHM}\sim$1\,arcsecond; $\sigma_{1.3\mathrm{\,mm}}\sim0.17$\,mJy\,beam$^{-1}$) allow us to classify the sample into individual sources, multiple systems, and potential lenses and/or close mergers. Interestingly, even at this high angular resolution, 73\,per\,cent of our detections are single systems, suggesting that most of these galaxies are isolated and/or post-merger galaxies. For the remaining detections, 20\,per\,cent are classified as multiple systems, 5\,per\,cent as lenses and/or mergers, and 2\,per\,cent as low-$z$ galaxies or Active Galactic Nuclei. Combining the \textit{Herschel}/SPIRE and ALMA photometry, these galaxies are found to be extreme and massive systems with a median star formation rate of $\sim$ 1,500\,$\mathrm{M_{\odot} yr^{-1}}$ and molecular gas mass of $M_{\mathrm{gas}}\sim10^{11}$\,$\mathrm{M_{\odot}}$. The median redshift of individual sources is $z\approx2.8$, while the likely lensed systems are at $z\approx3.3$, with redshift distributions extending to $z\sim6$. 
Our results suggest a common star-formation mode for extreme galaxies across cosmic time, likely triggered by close interactions or disk-instabilities, and with short depletion times consistent with the starburst-type population. Moreover, all galaxies with $S_{\mathrm{1.3mm}}\geq13$\,mJy are gravitationally amplified which, similar to the established $S_{500\mathrm{ \mu m}}>100$\,mJy threshold, can be used as a simple criterion to identify gravitationally lensed galaxies.

\end{abstract}




\section{Introduction}
\par Early observations conducted by the \textit{Infrared Astronomical Satellite (IRAS)}, even limited to the low-redshift universe ($z\lesssim0.1$), managed to show that astronomical objects emit an amount of energy in the infrared (IR) and sub-millimeter (submm) regime similar (or even higher) to that observed at ultraviolet (UV) and  optical wavelengths. 
This result made evident the potential biases of surveys of the star formation activity in the universe due to the presence of embedded stars in areas with great abundances of dust and gas \citep[][]{Casey2014}. Later, a population of high redshift galaxies whose infrared emission is dominated by dust heated by young stars, that contribute significantly to the IR and submm emission in the Universe, was confirmed. These sources, first found in maps of a few\, square\,arcminutes, were called Submillimeter Galaxies (SMG) \citep[e.g.][]{Smail_1997, Hughes1998}{}{}.

\par The SMG are characterized by having infrared luminosities ($L_{\mathrm{IR}}$) of $\gtrsim 10^{12}$\,L$_{\odot}$, equivalent to those of Ultra-luminous infrared galaxies (ULIRGs) and even reaching the extreme luminosities of Hyper-luminous infrared galaxies (HyLIRGs, $L_{\mathrm{IR}}\gtrsim 10^{13}$\,L$_{\odot}$). These luminosities correspond to star formation rates (SFR) of $\gtrsim 100-1,000$\,M$_{\odot}$\,yr$^{-1}$. Some spectroscopic studies have probed the CO line emission in this population, finding, typically, large reservoirs of molecular gas ($M_{\mathrm{gas}}>10^{10}\,\mathrm{M_\odot}$) distributed in a varied morphology that includes disk galaxies and compact star formation bursts of gas-rich galaxies in interaction \citep[e.g.][]{Oteo_2016}{}{}. The inferred depletion times from these observations are at the order of tens to a few hundred Myrs, which is short compared to the $\sim$1\,Gyr depletion time in normal galaxies \citep[e.g.][]{tacconi2010high, Saintonge_2011}{}{}. This difference opens the question of what is triggering the SFR in these galaxies, the high content of molecular gas or physical processes that are different in the early universe \citep[see reviews by][]{Blain2002,Carilli_2013, Casey2014}{}{}. 

\par Initially, studies such as \cite{chapman_2005} used optical spectroscopic observations to characterize the redshift distribution of 73 SMG, which had a median of $z=2.2$. However, the spectroscopy of these sources required very long integration times and that placed a limit on how many galaxies could be studied. Moreover, these relatively small samples were considerably affected by biases at the highest redshifts ($z>4$) introduced by the lack of detections at radio wavelengths, which at the time led to not having a good estimate of their position for spectroscopic follow up. Studies such as \cite{Aretxaga_2005,Aretxaga_2007} and \cite{Yun_2012}, using multi-wavelength photometry in different fields such as Lockman Hole East (LH), SXDF and GOODS-S, also found that the redshift distribution of this population of galaxies has a median value between 2 and 3, with a small deviation between different works. However, these studies were carried out with dozens or at most just over a hundred galaxies in areas of \,square\,arcminutes. More recent studies using medium-to-large sample sizes of up to hundreds SMG \citep[e.g.][]{da2015alma, Miettinen,Dudze_2020}, found similar redshift distributions with median values between 2 and 3. 
The identification of significantly larger samples of extreme SMG at high redshift requires of surveys over larger areas of hundreds of square\,degrees.

\par The \textit{Herschel} Astrophysical Terahertz Large Area Survey (\textit{H}-ATLAS) is one of the largest projects in time and spatial coverage that was carried out with the \textit{Herschel Space Observatory (Herschel)}, covering $\sim $600 square degrees of the sky towards the Galaxy And Mass Assembly (GAMA), North Galactic Pole (NGP) and South Galactic Pole (SGP) fields.
This key project includes continuum observations with PACS (at 100\,$\mu$m and 160\,$\mu$m) and SPIRE (at 250\,$\mu$m, 350\,$\mu$m and 500\,$\mu$m) \citep{Eales_2010}. 
Galaxies with flux densities increasing from 250\,$\mu$m to 500\,$\mu$m ($S_{\mathrm{250 \mu m}}<S_{\mathrm{350 \mu m}}<S_{\mathrm{500 \mu m}}$) can be associated with SMG at $z>2$\footnote{It must be noted, however, that low redshift galaxies with unusually cold temperatures can show a similar shape of their FIR spectrum.} \citep[][]{Hughes_2002}{}{}. These galaxies have been called \textit{``500\,$\mu$m risers"} \citep[][]{pope_2010,Cox_2011}, and are also known as red-\textit{Herschel} sources.

\par Using the first data release of the \textit{H}-ATLAS collaboration, \cite{Ivison_2016} identified a sample of 7,961 red sources detected above the 3.5$\sigma $ threshold at 500\,$\mu$m and with flux ratios of $S_{500\,\mu \mathrm{m}}$/$S_{250\,\mu \mathrm{m}} \geq 1.5$ and $S_{\,500 \mu \mathrm{m}}$/$S_{350\,\mu \mathrm{m}} \geq 0.85$. 
After a visual inspections of 2725 galaxies, a subsample of $\sim100$ was drawn to be targeted with the James Clerk Maxwell Telescope and the Atacama Pathfinder Experiment. 
Subsequent studies and follow up observations have focused on these galaxies and have allowed to spectroscopically confirm and study some of the most distant and most extreme SMG that are known to date \citep[e.g.][]{Ivison_2016,Fudamoto_2017,Zavala,Ma_2019,bakx2020iram, Montaña_2021}.

\par  Observations with Atacama Large Millimeter/submillimeter Array (ALMA) and \textit{James Webb Space Observatory (JWST)} at resolutions less than 1\,arcsecond have shown that, what we considered as extreme red galaxies based on  \textit{Herschel} observations, in many cases are galaxies that are being gravitationally amplified or multiple galaxies blended within the \textit{Herschel} large beam. Some of these galaxies and galaxy groups might be interacting galaxies tracing overdense environments, which could be progenitors of elliptical galaxies or the massive galaxy clusters in the local Universe \citep[e.g.][]{Fudamoto_2017,Oteo_2018,Amvrosiadis_2018, Zavala, Zavala_2019, XiaAn_2019,Jones_2023}{}{}.

\par In this work we compile the currently largest sample of red-\textit{Herschel} galaxies from the {\it H}-ATLAS with publicly available $\sim$1\,arcsecond angular resolutions ALMA observations. The high angular resolution data allow us to quantify the relative abundance of single, multiple, and potentially lensed galaxies within this population. Using the SPIRE and ALMA photometry, we further characterize the redshift distribution and physical properties of those red-{\it Herschel} sources identified as individual galaxies and potentially lensed or close merging systems. The analysis of the {\it Herschel} sources in our sample identified as multiple systems (i.e. blending two or more ALMA detections) will be presented in a forthcoming paper.

This paper is structured as follows: Section \ref{dataandsample} provides a detailed description of our sample, the Herschel-Alma Red sources Public Archive Study (HARPAS), as well as of the data processing and source detection; Section \ref{sec:analysisandresults} presents the methodology and results; Section \ref{sec:discution} compares our results in the framework of the different studies of red-\textit{Herschel} galaxies as in the general scheme of the SMG population; finally, we summarize our results and conclusions in Section \ref{sec:summary}.
\par In this work we assume a $\Lambda$CDM cosmology using ${H}_{0}=70\,\mathrm{km}\,{{\rm{s}}}^{-1}\,{\mathrm{Mpc}}^{-1}$, ${{\rm{\Omega }}}_{{\rm{M}}}=0.3$ and  ${{\rm{\Omega }}}_{{\rm{\Lambda }}}=0.7$.

\section{Data and sample selection}
\label{dataandsample}

\subsection{Red \textit{H}-ATLAS galaxies observed with ALMA}
We build a sample of red-\textit{Herschel} galaxies (defined with a simple selection criterion of $S_{\mathrm{250 \mu m}}<S_{\mathrm{350 \mu m}}<S_{\mathrm{500 \mu m}}$) using the data release catalogs from \textit{H}-ATLAS \citep[][]{Valiante_2016, Maddox_2018}{}{}. We find 6,194 sources that meet the criterion, representing only 1.4\,per\,cent of all the \textit{H}-ATLAS detections. 
\par We crossmatch our catalog with public data from the ALMA archive using Python Virtual Observatory (PyVo) v1.5.1 and adopting a 5\,arcseconds search radius. We find that 3,187 of these red-\textit{Herschel} sources have been observed in 41 public ALMA projects, with 3,257 pointings in band 6, 192 in band 3, 96 in band 4, 18 en band 5, 61 in band 7 and 20 in band 8.  Band 6 observations represent 97\,per\,cent (3,089) of the total red \textit{Herschel}-ATLAS sources observed with ALMA\footnote{Projects 2016.1.00087.S (P.I. S. Chapman), 2018.1.00489.S and 2018.1.00526.S (P.I. I. Oteo). The Science Verification Project 2011.0.00017.SV (P.I ALMA Observatory) includes one additional red-\textit{Herschel} source. This project, however, does not have available data in the NRAO Data Archive.}. 
We thus focus our analysis on this Band 6 data-set, which benefits from an homogeneous depth and angular resolution.

\subsection{ALMA data \label{sec:ALMAdata}}
To extract the available data at 1.3\,mm (233\,GHz), we first obtained the calibrated visibilities or measurement sets (MS) from the NRAO Data Archive\footnote{\url{https://data.nrao.edu/portal/}}. The standard reduction and calibration of the ALMA pipeline was adopted. Projects such as 2018.1.00526.S (P.I. I. Oteo) take advantage of the ALMA clustering algorithm, which allows for very short observations of less than 1 minute per source thanks to the sharing of calibration observations. This strategy is, however, limited to sources within 10 degrees (or within 1 degree in the case of long-baseline configurations or high-frequency observations) and may not be applicable to other samples.

 \par We then produced the images from the ALMA visibilities using \texttt{CASA} \citep[][]{TheCASATeam_2022}{}{} and the task \texttt{tclean}, which computes the Fourier transform of the \textit{uv} visibilities. In the case of fields with multiple observations, we combine all the available MS files into one single map.  
 First, we create the dirty maps in which we use \texttt{Sigma clip} from Astropy v5.3.1 to obtain a global value of the noise r.m.s in the maps. The sigma clipping method removes all the data that have values outside a given threshold, in our case $\pm 5\sigma$. 
 
\par
\begin{figure}
    \centering
    \includegraphics[width=\columnwidth]{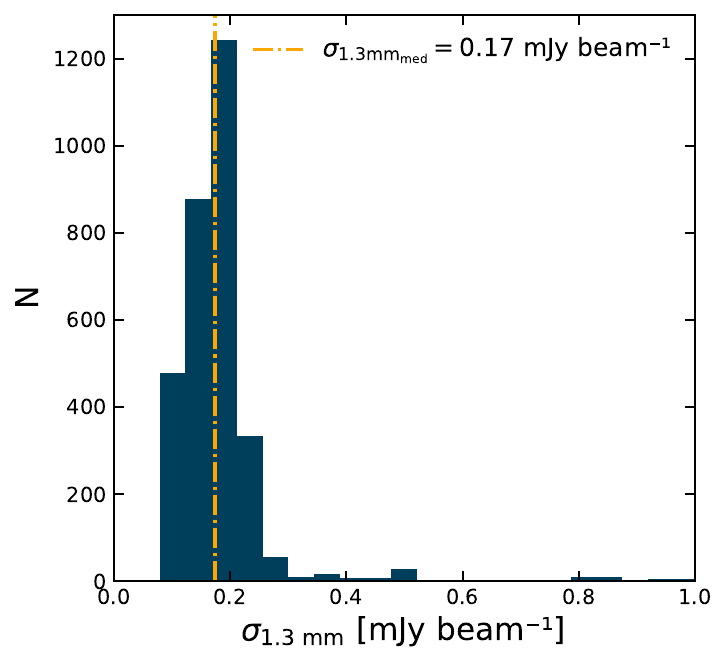}
    \caption{Histogram of the r.m.s. noise of the 1.3\,mm ALMA (excluding 8 fields with r.m.s. greater than 1\,mJy\,beam$^{-1}$ identified as AGN). The median r.m.s. of the sample is 0.17\,mJy\,beam$^{-1}$. Note that the expected 1.3 mm flux density of a red-\textit{Herschel} source with $S_{500 \mu \rm{ m}} \sim 40$\,mJy (i.e. the \textit{H}-ATLAS catalog detection limit) is 1.5\,mJy, extrapolated using a modified black body at $z = 2$ and with typical SMG values with luminosity-weighted dust temperature $T_{\rm dust}$ = 30\,K, emissivity index $\beta = 1.8$, and $L_{\rm IR} = 6\times10^{13}$\,L$_\odot$.}
    \label{fig:sigma}
\end{figure}

 As a second step for the imaging, we use the \texttt{tclean} task to produce clean maps of each field with a clean threshold of  3$\sigma$ (being 1$\sigma$ the r.m.s. value of the sigma-clipped dirty map described above). The noise values in the clean maps range from 0.1 to 3.6\,mJy\,beam$^{-1}$ with median of 0.17\,mJy\,beam$^{-1}$. Figure \ref{fig:sigma} shows the distribution of r.m.s. values for those fields with r.m.s. less than 1\,mJy. The eight fields with r.m.s. greater than 1\,mJy\,beam$^{-1}$ were identified as being centered on Active Galactic Nuclei (AGN) in Section \ref{sec:samplecleaning}. The extremely bright AGN (SNR $>100$) detected in these fields limit the dynamical range of the observations introducing artifacts in the maps which increase the measured r.m.s. All the maps have a size of $216 \times 216$\,pixels, with pixel sizes of 0.14--0.25\,arcseconds, and are cut to a primary beam response of 0.2. The beam-sizes range from 0.8 to 1.6\,arcseconds and have a mean value of $1$\,arcsecond. We used a \texttt{natural} weighting, which uses \textit{uv} visibilities with alike data weights in the weight grid following the sample density pattern on the \textit{uv}-plane, which increases the signal to noise at the expense of a slight lower angular resolution \citep{cycle7technical}. All the pointings together cover an area of $\sim$ 1,000\,square\,arcminutes, larger than most of the ALMA blind surveys to search for Dusty Star-Forming Galaxies (DSFG) \citep[e.g.][]{Zavala_2021,franco2023unveiling}.

\subsubsection{Source Detection and False Detection Rate}
To identify sources in the ALMA data, we make use of the maps that are not corrected by the primary beam response created as described in Section \ref{sec:ALMAdata}. These maps have the advantage of having a constant noise, which can be used to easily create signal-to-noise ratio maps. 

\par The detection is performed by searching the maps for pixels that have values above a certain threshold within a radius of 16.6 arcseconds to avoid the noisiest regions at the edges of the map. Within this region, the primary beam response is above 0.3. When a pixel exceeding this value is identified, we mask a region equal to 1.5 times the beam-size. This process is done iteratively until no more pixels are found over this threshold in a given map.

\par
\begin{figure}
    \centering
    \includegraphics[width=\columnwidth]{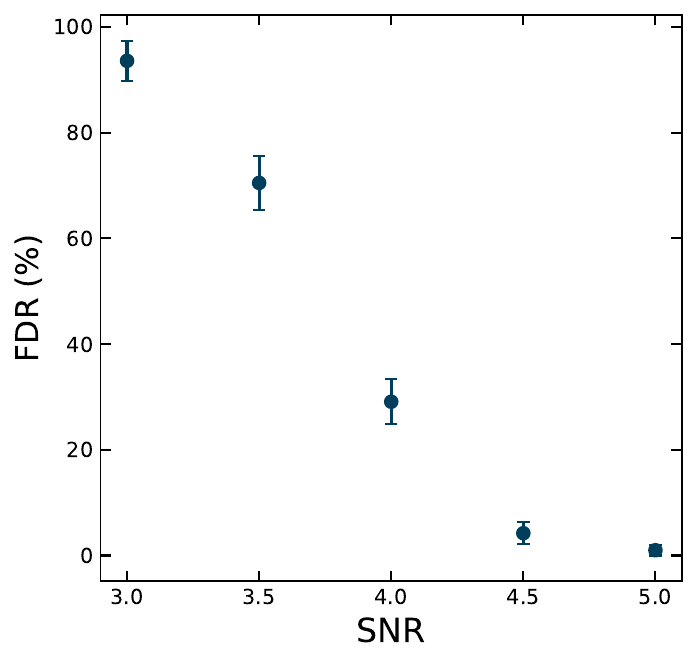}
    \caption{False detection rate (FDR) as a function of SNR, estimate within a 16.6\,arcseconds radius to avoid the noisier regions at the edge of the 1.3 mm ALMA maps. We adopt a 5$\sigma$ detection threshold where the FDR is estimated to be $<1$\,per\,cent. The uncertainties of the FDR were estimated through a bootstrap technique with 100 realizations with different random subsamples.}
    \label{fig:FDR}
\end{figure}

 To calculate the false detection rate (FDR) as a function of signal-to-noise ratio, we first run our detection algorithm with a minimum threshold of 3$\sigma$ and define these detections as "Positive" ($D_P$). Then, we multiply the maps by $-1$ and run again the detection algorithm, defining such detections as "False" ($D_F$) since they are expected to arise only from noise fluctuations. Finally, the FDR is estimated as: 
\begin{equation}
    \mathrm{FDR} [\%]= \Bigg(\dfrac{D_F}{D_P}\Bigg)\times 100.
\end{equation}
This procedure was carried out for different thresholds, as shown in  Fig. \ref{fig:FDR}. As it can be seen, at $>5\sigma$, the FDR drops below $\sim1$ per cent. Therefore, to be conservative, in this work we use a 5$\sigma$ detection threshold. We note that the main results do not change significantly if we adopt a different threshold (e.g. 4.5$\sigma$, where the FDR is around 5 per cent).

\par Using this 5$\sigma$ threshold, we identify 673 fields with non detections (same number at 4.5$\sigma$) and 2,416 fields with at least one detection. Out of these 2,416 fields, we found multiplicity in 494 maps, which means that these fields have at least two possible sources detected over 5$\sigma$ in an area similar to the \textit{Herschel}/SPIRE beam at 500\,$\mu$m. 

\subsubsection{Flux density measurements}
\par For each detected source, we measure the flux density at the brightest pixel and the corresponding value within a given aperture. In the first case, the measured flux density (i.e. brightest pixel value) and its associated uncertainty ($\sigma_p$, the standard deviation of the constant-noise map) are corrected by the primary beam response at the position of the detection. For the aperture measurements, we coadd the flux density within different apertures (from 1.5 to 3 times the beam-size) and correct them dividing by the synthesized beam. We adopt the flux density value where its aperture growth curve becomes constant. The uncertainty of the aperture flux density ($\sigma_a$) was estimated following equation 2 of \cite{Bethermin_errorapertura}. After all the corrections were done, we added the calibration uncertainty ($\sigma_{ca}$) for ALMA, that is a around 6 to 10\,per\,cent of the flux density measurement \citep[][]{cycle7technical}{}. The final value for the uncertainty is given by,

\begin{equation}
    \sigma_c=\Big(\sigma_{ca}^2+\sigma_{x}^2\Big)^{1/2},
    \label{eq:cuadratura}
\end{equation}
using $\sigma_{x}=$ as $\sigma_p$ or $\sigma_a$. We adopt $\sigma_{ca}=0.1$.

\subsection{Sample cleaning \label{sec:samplecleaning}}
\par We cleaned our sample of contamination from AGN and low-redshift galaxies ($z<1$) using the The Million Quasars (MILLIQUAS) Catalog \citep[]{Milliquas} and the 17th Data Release of the Sloan Digital Sky Survey (SDSS) \citep[]{SDSS_DR17}. 

Any ALMA detection lying at $\leq 2\,$ arcsecond from a source in the MILLIQUAS and SDSS catalogs is considered a match and is therefore discarded from our analysis. We classified 43 ALMA sources as AGN or galaxies at $z<1$. These sources are presented in Table \ref{tab:agn_lowz}.

\subsection{Flux density estimation in SPIRE maps \label{sec:deblending}}
Taking advantage of the better positional accuracy from the ALMA observations, we re-estimate the SPIRE fluxes in the three different bands using a deblending technique similar to the one presented by \cite{Michal+2017}.

\par First, we made 120x120\,arcseconds$^2$ postage stamps from the background subtracted and unfiltered SPIRE maps for each source in our sample, centered at the center of the ALMA maps. Second, to account for the emission from any nearby galaxy around the ALMA position, we searched for 250\,$\mu$m-detected sources (using a 4$\sigma$ threshold) outside of our ALMA detection radius (16.6 arcsecond). Third, we performed a simultaneous 2D Gaussian fit for each source detected in the postage stamp, including the ALMA detection inside our detection radius and any 250\,$\mu$m source outside this radius.

During the fitting, we fix the FWHM of the Gaussian to the FWHM measured from the Neptune PSF maps and fix the position to that from the ALMA and the 250\,$\mu$m maps. PLUM - Fields are also treated as single sources since the sources found are separated by less than $3$\,arcsecond, which is smaller than the {\it Herschel}/SPIRE pixels ($6\times6$, $8\times8$, and $12\times12$ arcseconds$^{2}$ at 250, 350, and 500 $\mu$m, respectively). In this case, we adopt the position of the brightest component, but we get similar results if the centroid of the sources is adopted.  
The Gaussian amplitude and the background level are set as free parameters. The fitted Gaussian amplitude represents the final deblended flux density of the source. This process is repeated for each of the three bands, but the search of sources in the outer regions is always based on the 250\,$\mu$m map, which has the best angular resolution. The uncertainties in the fitted amplitudes are used as the flux density errors, which are added in quadrature along with the confusion noise and the absolute calibration errors. The SPIRE maps have a confusion noise of 5.8, 6.3 and 6.8\,mJy/beam, respectively and a calibration uncertainty of 5.5 per cent \citep[][]{Valiante_2016}{}{}.

\par
\begin{figure*}
    \centering
    \includegraphics[width=\textwidth]{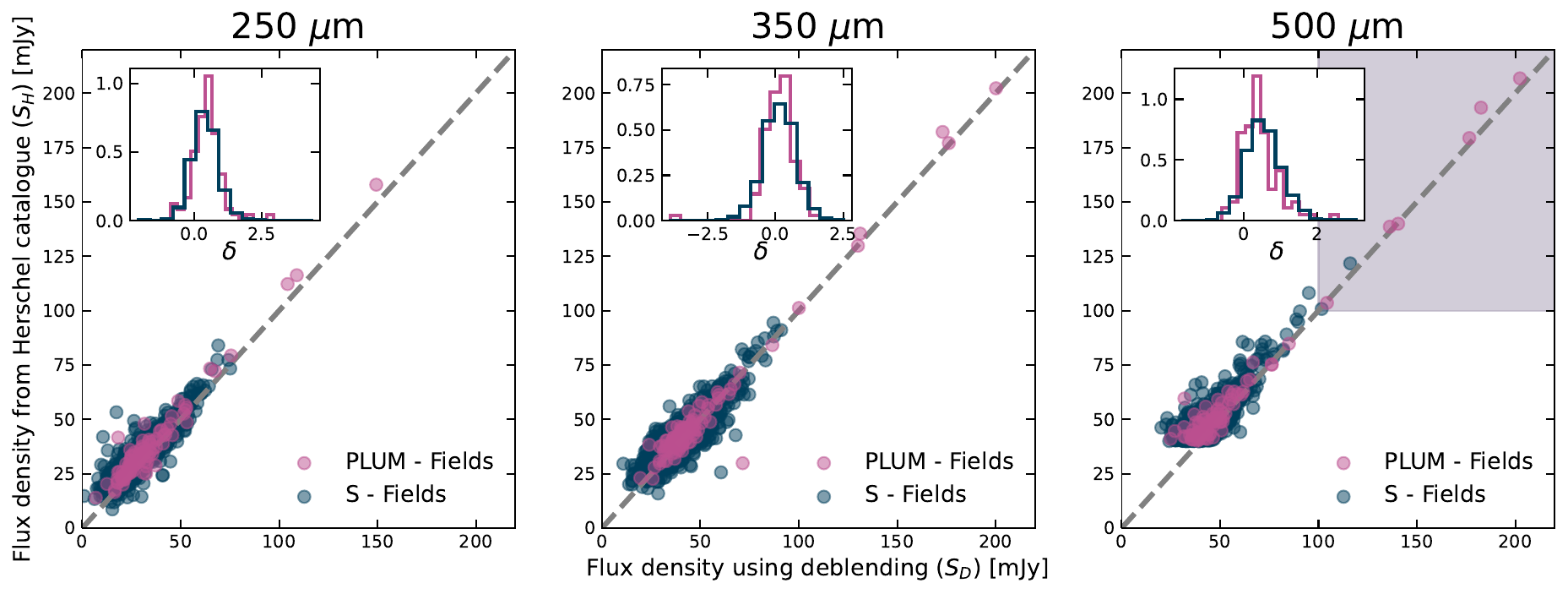}
    \caption{Comparison between the SPIRE fluxes from the \textit{H}-ATLAS catalog \citep[][]{Valiante_2016, Maddox_2018, Ward_2021}{}{} and the optimized fluxes obtained in this study. Single sources (S - Field) are shown in blue, while the potential lensed and/or unidentified mergers (PLUM - Fields) are shown in pink. Our fluxes were obtained through a deblending technique (Section \ref{sec:deblending}), in which a two-dimensional Gaussian fit was made, fixing the position of the source detected in ALMA and the FWHM to the resolution of the SPIRE map. The shaded region indicates the sources that have fluxes over 100\,mJy at 500\,$\mu$m in both catalogs. Sources above this limit are expected to be gravitationally amplified systems \citep[][]{negrello_2010}{}{}, as suggested also by ALMA observations and their PLUM classification (with the exception of two single sources). The inset histograms show a similar scatter of the flux density measurements (in terms of $\delta = (S_H - S_D)/\sigma_D$) for both S - Fields and PLUM - Fields.}
    \label{fig:deb}
\end{figure*}

 Our deblending flux densities are compared to the  ones from the \textit{H}-ATLAS catalogs in Fig. \ref{fig:deb}. Overall, there is a very good agreement between the two sets of values, with a small dispersion. The largest difference is at the faint-end of the 500\,$\mu$m flux densities, where our estimations go below the $\sim$40\,mJy detection limit of \textit{H}-ATLAS \citep{Valiante_2016}. Flux densities at 500\,$\mu$m from the {\it H}-ATLAS catalog are $\sim10$\,per cent higher than our deblended estimates. If, for a more conservative comparison, only those sources with deblended flux densities above the \textit{H}-ATLAS detection limit are considered, the 500\,$\mu$m flux densities from the \textit{H}-ATLAS catalogues are $\sim$ 6 and 8\,per cent higher than our deblended flux densities for the PLUM - Fields and S - Fields, respectively. This difference may change the number of galaxies classified as red-\textit{Herschel} sources in the sample and have implications in the estimations of redshifts and other properties, as discussed in Section \ref{sec:discution}.

\section{Analysis and results}
\label{sec:analysisandresults}

\subsection{Classification}
\label{sec:classification}
\par
\begin{figure}
    \centering
    \includegraphics[width=\columnwidth]{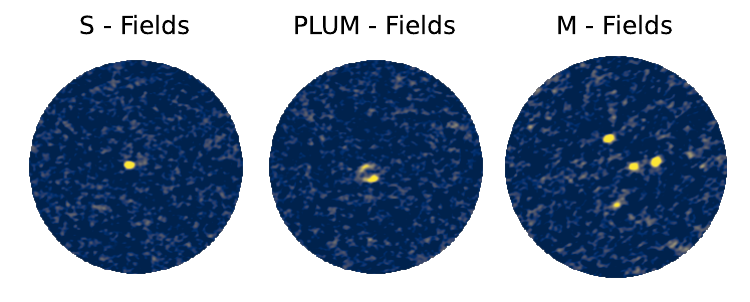}
    \caption{Examples of 1.3\,mm ALMA maps for the three subsamples: S - Fields (singles; HATLASJ232147.3-340644 with one detection), PLUM - Fields (potential lenses and unidentified mergers; HATLASJ234350.6-295817 with two arc-like detections) and M - Fields (multiples; HATLAS	J122459.1-005647 with four detections). The values in parentheses indicate the number of fields in each classification, out of the 2416 ALMA maps with detections (i.e. excluding 43 fields associated to AGN or nearby galaxies, and the 673 fields with no ALMA detections above $5\sigma$). These maps have a radius of $\sim 18.6$\,arseconds and a median r.m.s. of 0.17\,mJy beam$^{-1}$. }
    \label{fig:sample_alma}
\end{figure}

 Fields with at least one source detected above the adopted threshold (i.e. 2416 fields) were classified in three different groups depending on the number of detections and their proximity. Examples of the three classifications, which are described in detail below, can be found in Fig. \ref{fig:sample_alma}. Additionally, fields with no detections are classified as N - Fields and are briefly discussed in Section \ref{N-fiels}.

\subsubsection{S - Fields}
Fields that have only a single detection above 5$\sigma$ within the 16.6\,arcseconds search radius are classified as S - Fields. These are the dominant class with a total of 1,762 sources (1,761 at 4.5$\sigma$). The flux density range is 0.78\,mJy to 14.74\,mJy with a median of 3.14\,mJy at 1.3\,mm.

\subsubsection{PLUM - Fields}\label{sec:PLUMS}
Potential Lenses and Unidentified Mergers (PLUM - Fields) are fields with at least two detections separated by no more than 3\,arcseconds, which may be lensed systems or closely interacting galaxies. The 3\,arcsecond separation was chosen based on the typical separation of strong lenses reported in the literature \citep[e.g.][]{Zavala,Gururajan,Bendo}{}. An additional visual inspection of the whole sample was performed to identify arc-like or extended structures (larger than the synthesized beam) that were originally identified as a single source by our source-detection algorithm, finding 46 such cases. 

\par In total, this category contains 137 fields (149 at $>4.5\sigma$) after cleaning for AGN and low-$z$ galaxies interlopers. The flux density range is 1.00\,mJy to 35.22\,mJy with a median of 4.75\,mJy at 1.3\,mm. From this subsample, we identified 12 fields that are robust lenses confirmed in the literature \citep[][]{Bussmann_2013, Negrello+2016,Fudamoto_2017, Lewis_2018, Majon+2019,urquhart+2022}{}{} or show a strong arc-like morphology in the ALMA observations (supporting their lensed nature; see further discussion in Section \ref{sec:fluxandcolors}). 

\subsubsection{M - Fields}
Multiple fields include those with two or more sources that are separated by more than 3 arcseconds. This category contains 474 systems. From these, 420 have only two components, 51 have three sources, and 3 fields have 4 detections.
The sources in these fields have an average flux density of $\sim$2.7\,mJy at 1.3\,mm. The analysis and physical properties of these fields will be presented in a forthcoming paper. 

\subsubsection{N - Fields}\label{N-fiels}
This classification corresponds to fields where no detections above a 5$\sigma$ level are found. There are 673 fields in this category with a median r.m.s. of 0.19\,mJy\,beam$^{-1}$. The distribution of r.m.s. values in these N - Fields is similar to those from the other categories where sources are detected. A fraction of these fields may therefore correspond to multiple systems of sources with flux densities below the 5$\sigma$ detections threshold, as it has been suggested in previous follow-up studies of red-Herschel sources \citep[e.g.][]{Greenslade_2020,Montaña_2021}.

\begin{table}
    \centering
    \begin{tabular}{ccc}
    \hline
   
      Median   &  S - Fields & PLUM - Fields\\
      \hline
      $z$  & $2.78\pm 0.03$ [2.13-3.43]& $3.28\pm0.12$ [2.53-4.09] \\
       $L\mathrm{_{IR} [10^{12}\,L_{\odot}]}$  & ($8.9\pm 0.2)$  [5.3-14] & (13.5 $\pm 1.2) $ [8.4-24] \\
       SFR$ [\mathrm{M_{\odot} yr^{-1}]}$ &$1300\pm30$ [800-2160] &$2000\pm 170$ [1240-3590] \\
       $M_{\mathrm{gas}}\mathrm{[10^{11}\,M_{\odot}]}$  & $(4.1\pm0.1)$ [2.6-6.2]&  $(5.7\pm 0.3)$ [3.8-9.7]\\
       $M_{ \mathrm{dust}}\mathrm{[10^{9}\,M_{\odot}]}$   & $(2.20\pm0.04)$ [1.5-3.2]&  (3.00 $\pm 0.20)$ [2.1-4.9]\\
          \hline
    \end{tabular}
    \caption{Median values of the physical properties estimated combining the deblended SPIRE/\textit{Herschel} and ALMA photometry for the S - Fields and PLUM - Fields samples. The uncertainties of the medians where calculated through a bootstrap of 100 synthetic subsamples. The values inside the square brackets indicate the limits containing the 68 per cent of the distributions (i.e. the 16th/84th percentiles). 
    }
    \label{tab:results}
\end{table}

\subsection{Redshifts, Luminosities and Star Formation Rates\label{subsec:red_lum_sfr}}
\par We use the software package \texttt{Millimeter Photometric Redshift} \citep[MMPz,][]{Casey_2020} to estimate photometric redshifts and infrared luminosities for the S - Fields and PLUM - Fields samples, using the {\it Herschel} and ALMA photometry derived in this work. For PLUM - Fields, the total ALMA 1.3 mm flux density is estimated by adding the flux density from individual components with their uncertainties added in quadrature.

This software fits modified black bodies to photometric data-points and obtains the most probable combinations of $z_{\rm phot}$, $L_{\mathrm{IR}}$ and  $\lambda_{\mathrm{peak}}$. The best set of parameters is selected based on the empirical relationship between  $\lambda_{\mathrm{peak}}$ and $L_{\mathrm{IR}}$ \citep[][]{Casey_2018}:

\begin{equation}
\langle \lambda_{\mathrm{peak}}(L_{\mathrm{IR}}) \rangle  \ [\mathrm{\mu m}]=\lambda_0 \Bigg(\dfrac{L_{\mathrm{IR}}\ [\mathrm{L_{\odot}}]}{L_{t}}\Bigg)^{\eta},
\end{equation}
where $\lambda_0=102.8\ \pm\ 0.4\ \mu$m, $L_t=10^{12}$ L$_\odot$ and $\eta=-0.068\ \pm\ 0.001$.

\begin{figure}
    \centering
    \includegraphics[width=\columnwidth]{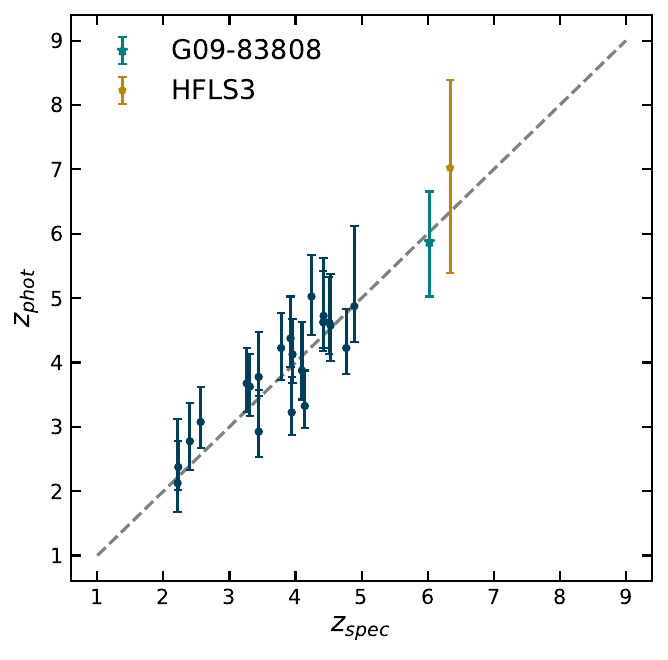}
    \caption{Comparison between the photometric redshifts obtained with MMPz in this work and spectroscopic determinations reported in the literature \citep[][and Montaña et al, in prep.]{Fudamoto_2017, bakx2020iram,Montaña_2021,urquhart+2022,Cox_2024}{}{}. Also included are the starburst galaxies HFLS3 \citep[$z=6.34$,][]{Riechers_2013}{}{} and G09-83808 \citep[$z=6.027$;][]{Zavala}{}{} with their photometric redshifts estimated using MMPz. Their median offset, $(z_{\mathrm{spec}}-z_{\mathrm{phot}} )/(1 + z_{\mathrm{spec}}) \approx 0.08$, confirms the reliability of our photometric redshifts to conduct a statistical characterization of the large sample of red-{\it Herschel} galaxies considered in this work.}
    \label{fig:compzspec_phot}
\end{figure}

To test the accuracy of this methodology, we compare the  photometric redshifts estimated for sources in our samples with available spectroscopic measurements in the literature (Fig. \ref{fig:compzspec_phot}).
The relative difference between photometric and spectroscopic redshifts $(z_{\mathrm{spec}}-z_{\mathrm{phot}} )/(1 + z_{\mathrm{spec}})$ is $\sim$8 per cent.
Thus, our photometric redshifts are accurate enough for a statistical characterization of this population of galaxies, the main goal of this paper. 

\par 

\begin{figure}
    \centering
    \includegraphics[width=\columnwidth]{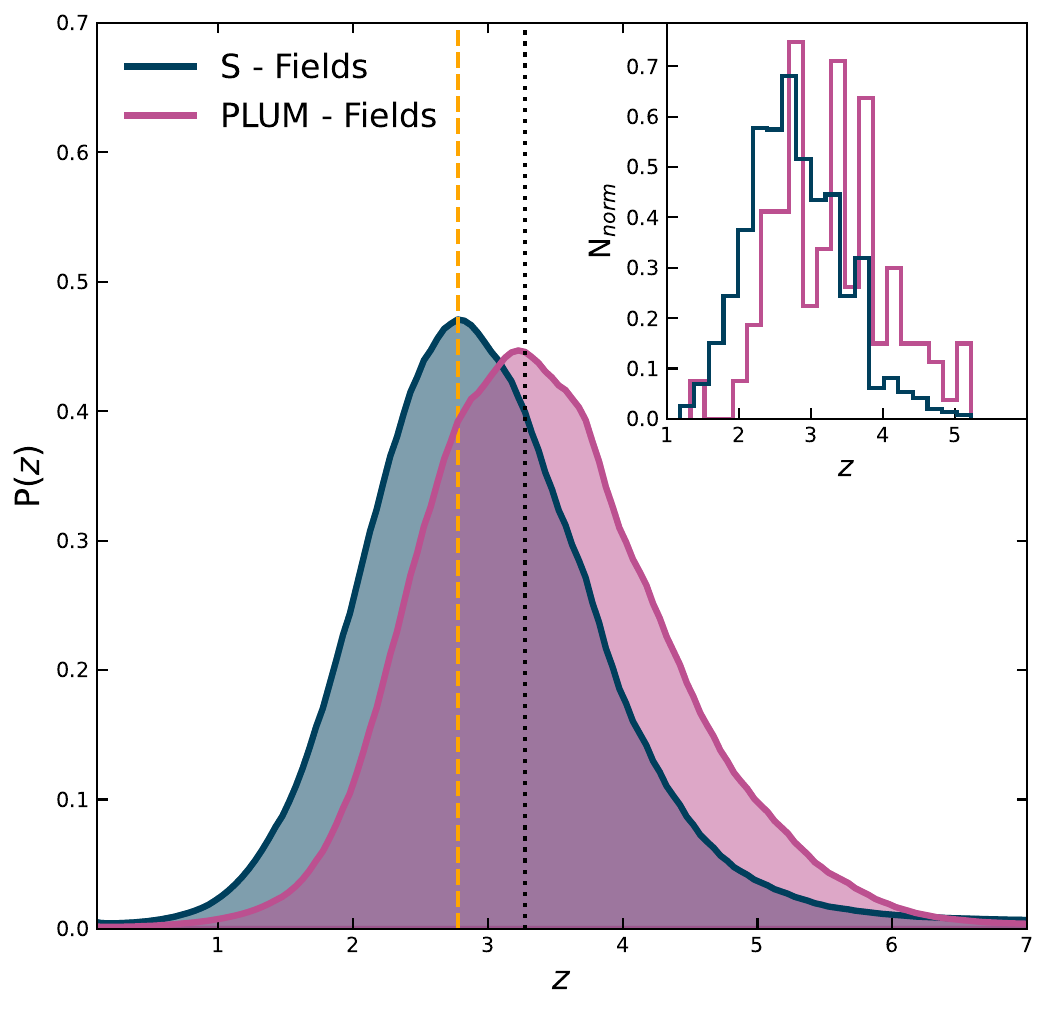}
    \caption{Redshift probability density distributions for S - Fields and PLUM - Fields (solid curves), and histograms of the best-fit photometric redshifts (inset). The median of the S - Fields is marked in yellow (dashed line) and the median for the PLUM - Fields is in black (dotted line). For the PLUM - Fields 60\,per\,cent of the sources are at $z>3$ while only 35\,per\,cent of the S - Fields lie at these higher redshifts.}
    \label{fig:z}
\end{figure}

Figure \ref{fig:z} shows the redshift probability distribution for the S - Fields and the PLUM - Fields, which have a median redshift of $2.78 \pm\ 0.03$ and $3.28 \pm\ 0.12$, respectively. The biased distribution towards higher redshifts of the PLUM-Fields may be reflecting the fact that the probability of strong gravitational lensing increases as a function of redshift (e.g. \citealt{Hezaveh_2011,Weiss_2013}). A deeper discussion and a comparison with other results from the literature is presented in Section \ref{secc:disc_zdist}.

\par
\begin{figure}
    \centering
    \includegraphics[width=\columnwidth]{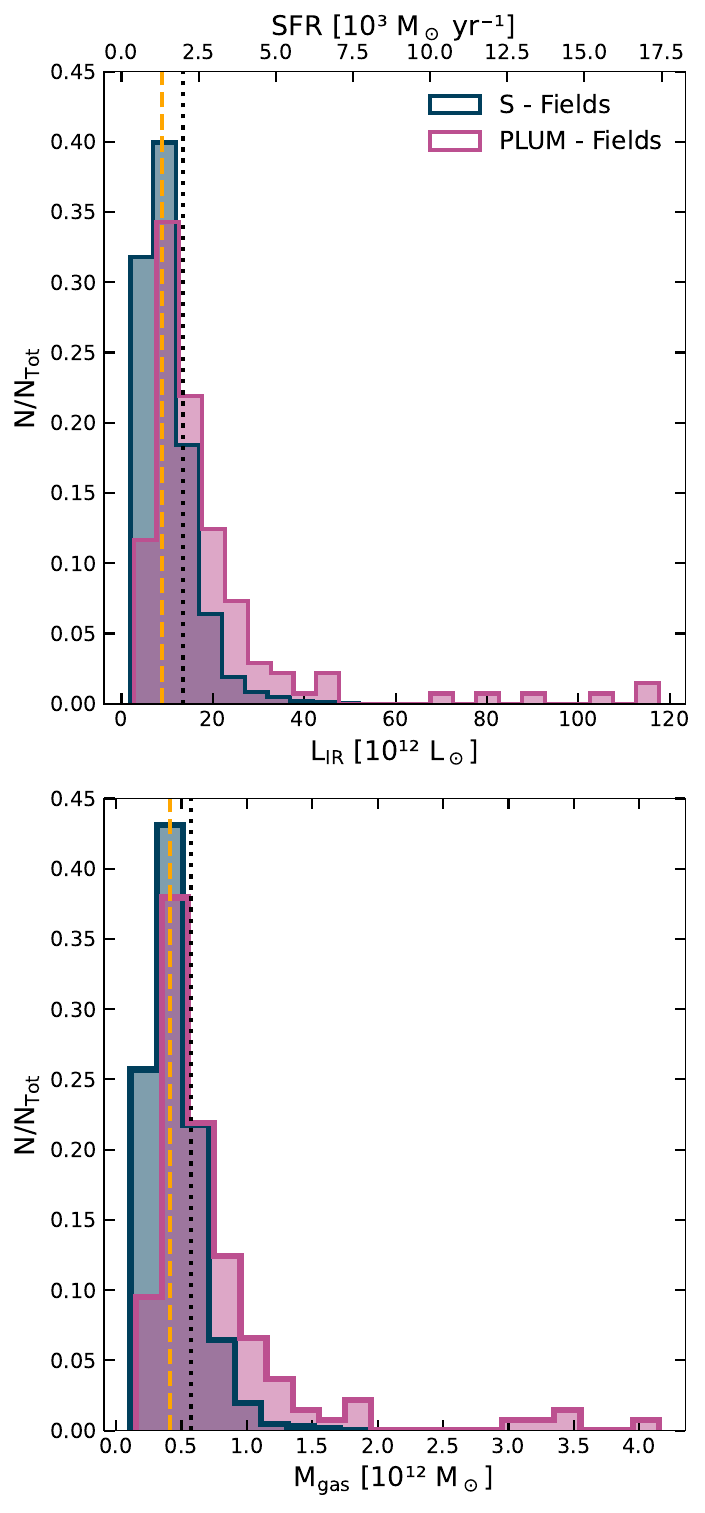}
    \caption{Histograms of the physical properties estimated for the S - Fields and PLUM - Fields samples using 1.3 mm ALMA and deblended \textit{Herschel}/SPIRE photometry: infrared luminosities and SFR (\textit{top}), and gas mass. The median values for the S - Fields are indicated with a yellow dashed line and those for the PLUM - Fields with a black dotted line.}
    \label{fig:prop}
\end{figure}

 The median $L_{\mathrm{IR}}$ of the S - Fields is $8.9\pm0.2\times 10^{12}$\,L$_\odot$ and for the PLUM - Fields is $13.5\pm1.2\times 10^{12}$\,L$_\odot$ (see Fig. \ref{fig:prop}). The larger $L_{\mathrm{IR}}$ values of the PLUM - Fields are expected since this sample includes those sources that are likely amplified by gravitational lensing effects (see, for example, Fig. \ref{fig:lensed}). Although most of the PLUM - Fields have extended detections with arc-like structures, it is also possible that a small fraction of these sources are interacting systems rather than lensed galaxies. If the higher luminosities of some of these sources were intrinsic, this would imply that galaxy interactions might trigger more extreme events of star formation, increasing (at least temporarily) their total IR luminosities.

\par Finally, we derive star formation rates from the IR luminosities following the relation presented in \citet[][]{kennicutt_2012},  $ \mathrm{SFR \,[M_{\odot}\,yr^{-1}]}= 1.48\times 10^{-10} \ L_{\mathrm{IR} }\mathrm{\,[L_{\odot}]}$, which assumes a \citet[][]{Kroupa_2001} IMF. We find a median SFR of $1300\pm30$\,$\mathrm{M_{\odot}\,yr ^{-1}}$ for the S - Fields and $2000\pm170$\,$\mathrm{M_{\odot}\,yr ^{-1}}$ for the PLUM - Fields. These SFR are also indicated in the top axis of the left panel of Fig. \ref{fig:prop} and compared with those from other galaxies in Section \ref{disc:LirandSFR}.

In an attempt to independently estimate the SFRs and gas masses (which are based on the ALMA photometry; Sec. \ref{sec:gas_masses}), we run MMPz using only the three {\it Herschel}/SPIRE data-points. This alternative, however, resulted in loose constraints on the SFR and, in some cases, overestimated redshifts, highlighting the importance of having photometric information probing the Rayleigh-Jeans regime. We therefore base our analysis and conclusions on the more robust and reliable results derived from the fit to the full dataset (i.e. using the {\it Herschel}/SPIRE and 1.3 mm ALMA photometry).

\subsection{Gas and Dust Masses}
\label{sec:gas_masses}
To infer the gas mass content of the galaxies in our sample, we use the relation given by \citet[][]{Scoville_2016}, which is calibrated using galaxies with CO line detections and Rayleigh Jeans dust continuum measurements of local star-forming galaxies, low-$z$ ULIGRs and $z\sim2$ SMG. This method has been previously used in similar studies of high-$z$ SMG \citep[e.g.][]{Harrington_2016, Birkin_2020, Castillo_2023}{}{}. This relation, which takes advantage of the fact that the Rayleigh Jeans regime is optically thin, is given by:
\begin{equation}
\begin{split}
 	M_{\mathrm{gas}} \ [\mathrm{M_{\odot}}	]=1.78\ S_{\nu_{\mathrm{obs}}}[\mathrm{mJy}](1+z)^{-4.8}\bigg(\dfrac{\nu_{850\mathrm{\mu \mathrm{m}}}}{\nu_{\mathrm{obs}}}\bigg)^{3.8} \\ \times (d_\mathrm{L} \rm{[Gpc]})^2 \bigg(\dfrac{6.17\times 10^{19}}{\alpha_{850}}\bigg) \dfrac{\Gamma_0}{\Gamma_{\mathrm{RJ}}}10^{10} \rm{M_{\odot}}, 
\end{split}
\end{equation}

where $\alpha_{850}=(6.17\pm1.7)\times 10^{19}$\,ergs\,s$^{-1}$\,Hz$^{-1}$\,M$_{\odot}^{-1}$ is a constant derived from the gas mass and the 850\,$\mu$m luminosity, $d_{\rm L}$ is the luminosity distance at the redshift $z$, $\Gamma_{\rm RJ}$ and $\Gamma_{0}$ are corrections due to the difference between the rest frame Planck function and the Rayleigh–Jeans at the source redshift and $z=0$, respectively. We assume a temperature of 25 K, a common mass-weighted value for the cold molecular gas, and a dust emissivity index $\beta = 1.8$.

\par We estimate a median $M_{\mathrm{gas}}$ of $(4.1\pm0.1)\times10^{11}$\,M$_{\odot}$ for the S - Fields and $(5.7\pm0.3)\times10^{11}$\,M$_{\odot}$ for the PLUM - Fields. Their distributions are shown in the bottom panel of Fig. \ref{fig:prop}. As a sanity check, we search for galaxies in our sample with CO measurements reported in the literature. For the 21 sources with available (mid-J) CO luminosities \citep[e.g.][]{Fudamoto_2017,Montaña_2021,Berta_2023,Hagimoto_2023}, we find a mean ratio of $\sim 0.8$ between the gas masses estimated in this work and those inferred from the CO line detections. Moreover, studies as \citet[][]{Zavala_2022}{}{} confirmed the reliability of this relation at higher redshifts based on observations towards a $z \approx 6$ gravitationally lensed dusty star-forming galaxy. Furthermore, \citet[][]{Castillo_2023}{}{} found that, when the same $\alpha_\mathrm{CO}$ value is adopted, the gas mass estimates using CO (1-0) and the \citet[][]{Scoville_2016}{} method agree within a factor of 2. 

 \par Dust masses are estimated using:
\begin{equation}
    M_{\rm dust} [\mathrm{M_{\odot}}]=\dfrac{S_{\rm obs}\, {d_\mathrm{L}}^2}{(1+z)k(\nu_{\rm obs})B(\nu_{\rm obs}, T_{\rm dust})}
    \label{eq:masapolvo},
\end{equation}
where $S_{\rm{obs}}$ is the observed flux density at 1.3\,mm, $d_{\rm L}$ is the luminosity distance, $z$ is the redshift (Section \ref{subsec:red_lum_sfr}) and $B(\nu_{\rm obs}, T_{\rm dust})$ is the Planck function evaluated at the observed frequency $\nu_{\rm obs}$ and a mass-weighted dust temperature $T_{\rm dust}$, for which we adopt a value of 25 K typical of high-$z$ SMG. The mass absorption coefficient $k(\nu_{\rm obs}) = k_0 (\nu_{\rm obs}/\nu_{850 {\rm \mu m}})^\beta$ is extrapolated from 850\,$\mu$m to 1.3 mm assuming $k_0=0.07\,\mathrm{m^2\,kg^{-1} }$ \citep[][]{James_2002}{}{} and $\beta = 1.8$. We find median dust masses of $(2.20\pm0.04)$$\times10^{9}$\,M$_{\odot}$ for the S - Fields and $(3.00\pm0.20)$$\times10^{9}$\,M$_{\odot}$ for the PLUM-Fields.

\section{Discussion}
\label{sec:discution}

\par In this section we discuss our results within the framework of studies based on smaller samples of red-\textit{Herschel} galaxies and the global context of the SMG population.

\subsection{Flux densities and colours\label{sec:fluxandcolors}}

\par
\begin{figure}
    \centering
    \includegraphics[width=\columnwidth]{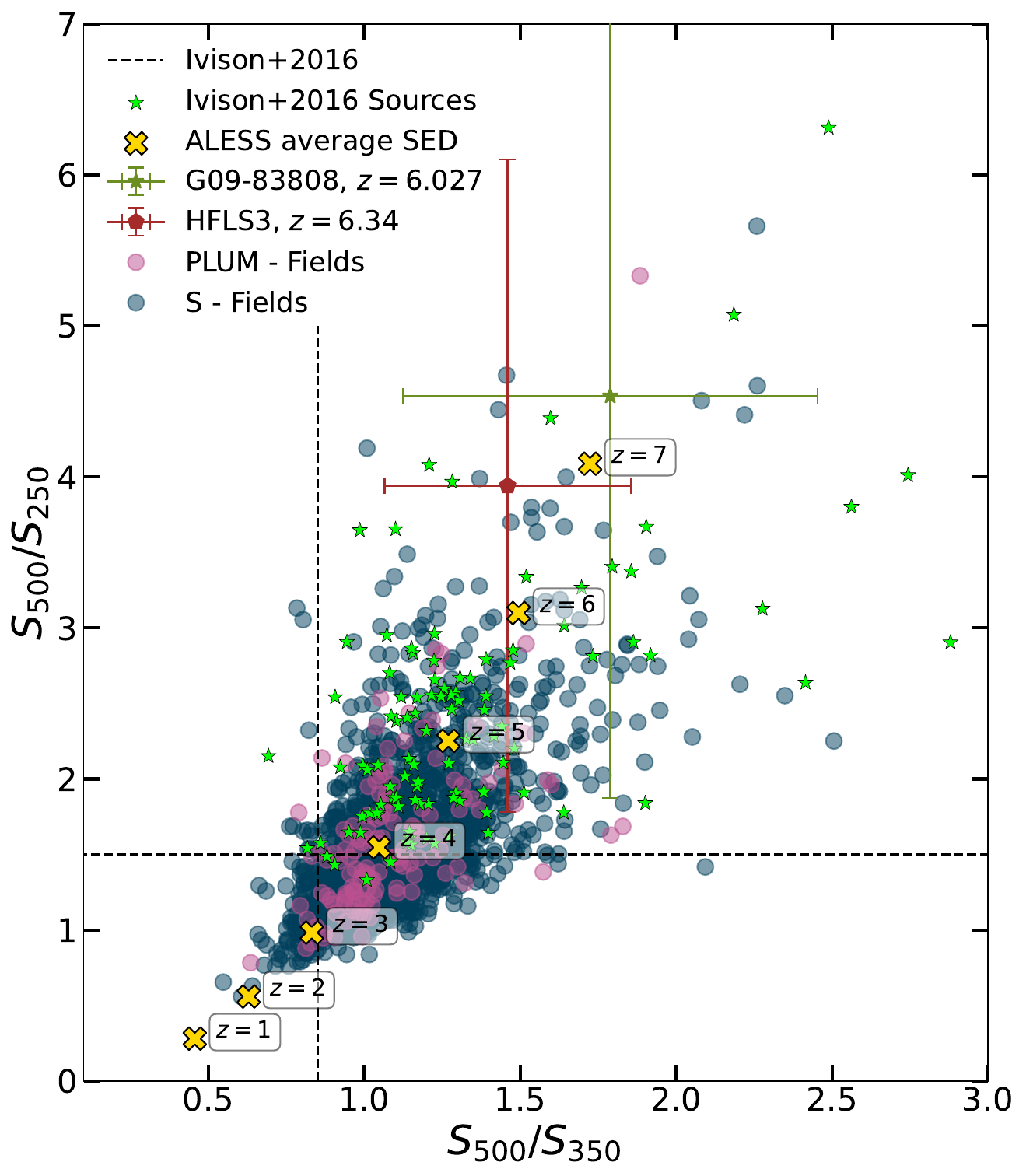}
     \caption{Colour-colour plot (after deblending) of the red-\textit{Herschel} sources in our samples. For comparison, the selection criteria from \citet[][]{Ivison_2016} are indicated by the dashed lines, and their sample of "ultra-red'' \textit{H}-ATLAS sources (using our deblending technique) are shown as green stars. We find that 44 per cent of the S - Fields and 51 per cent of the PLUM - Fields meet the \citet[][]{Ivison_2016} criteria. As a reference, the yellow \textit{x}-symbols indicate the redshift track of a source with a spectral energy distribution corresponding to the averaged template from ALESS \citep[][]{Cunha_2015_ALESS}{}{}. In addition, we show the position of two "ultra-red'' \textit{Herschel} sources with spectroscopic redshifts above 6 \citep[][]{Riechers_2013, Zavala}{}{}}
    \label{fig:color}
\end{figure}

 Figure \ref{fig:color} shows a colour-colour plot of our samples of red-\textit{Herschel} galaxies ($S_{\mathrm{250\mu m}}<S_{\mathrm{350\mu m}}<S_{\mathrm{500\mu m} }$), and compares it with the criteria from \citet[][i.e. $S_{ 500\,\mu \mathrm{m}}$/$S_{350\,\mu \mathrm{m}} \geq 0.85$ and $S_{500\,\mu \mathrm{m}}$/$S_{250\,\mu \mathrm{m}} \geq 1.5$.]{Ivison_2016}{}{} used to select ``ultrared'' $z \geq 4$ candidates \citep[e.g.][]{Ivison_2016,Dui_2018}{}{}. 
This figure shows that, after deblending, 94 per cent of the sources in our sample agree with $S_{ 500\,\mu \mathrm{m}}$/$S_{350\,\mu \mathrm{m}} \geq 0.85$, while only 45 per cent of the S - Fields and 52 per cent of the PLUM - Fields have $S_{500\,\mu \mathrm{m}}$/$S_{250\,\mu \mathrm{m}} \geq 1.5$. This implies that only 44 per cent of the S - Fields and 51 per cent of the PLUM - Fields meet both conditions. This might explain the relatively small fraction of $z>3$ sources in our sample --- i.e. the original sample selection includes sources that are not as ``red'' as expected for the highest redshift SMG (although they have similar high luminosities and SFR). This is relatively consistent with the redshifts estimated using MMPz, where 60 per cent of the PLUM - Fields are at $z>3$. 

\par
\begin{figure*}
    \centering
    \includegraphics[width=0.9\textwidth]{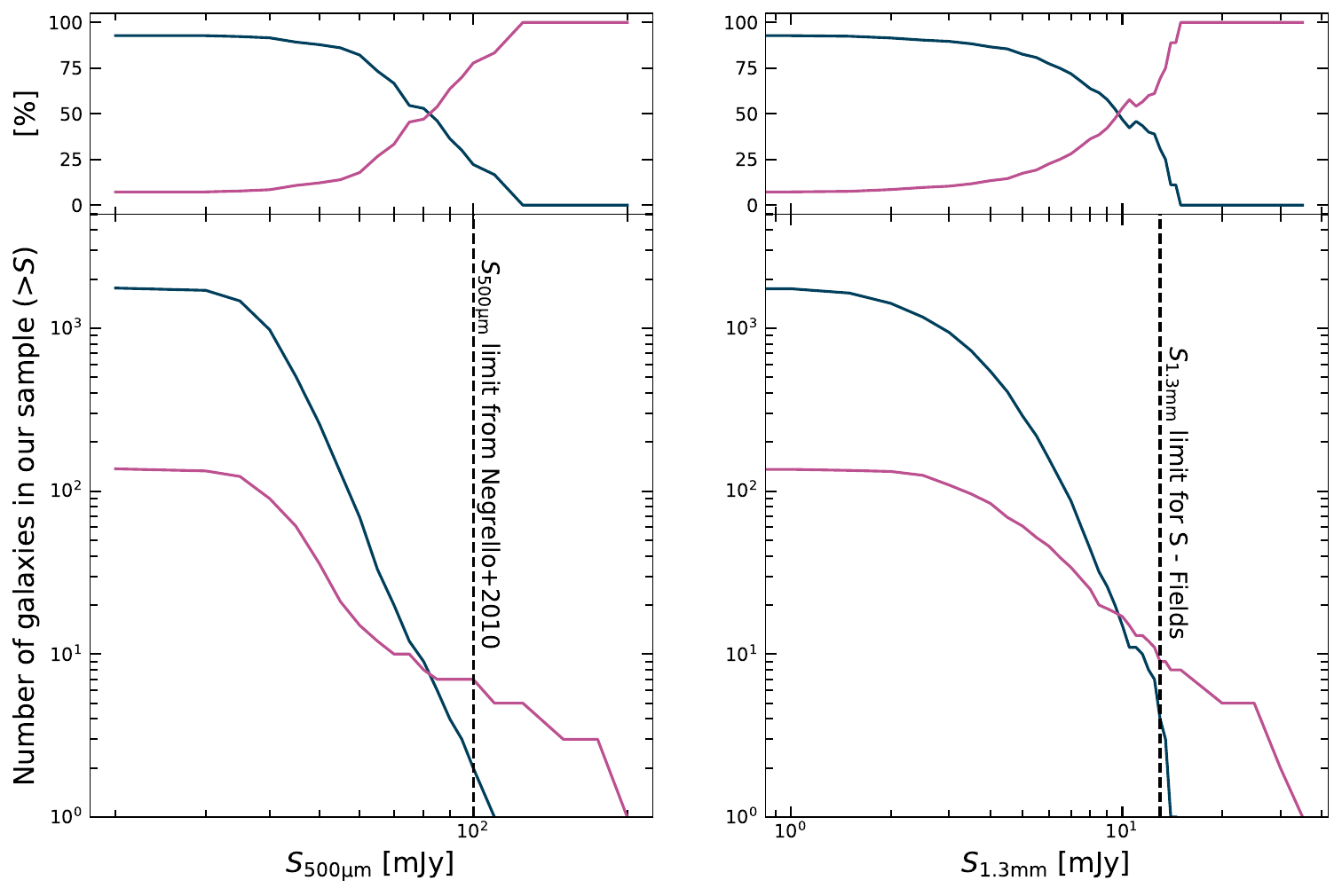}
    \caption{Cumulative number of galaxies in the S - Fields and PLUM - Fields of our sample (\textit{bottom panels}) and their fractional contribution (\textit{top panels}) as a function of $500 \mu {\rm m}$ ({\it left}) and 1.3\,mm ({\it right}) flux density. The $S_{1.3\rm mm}\geq 13.0$ is dominated by the PLUM - Fields with only 4 out of the 15 sources above this threshold being S-Fields. Nevertheless, all these 15 sources show evidence of gravitational lensing as discussed in the main text (see also Figure \ref{fig:lensed}). This 1.3 mm flux density limit can thus be used to identify DSFG amplified by strong  gravitational lensing effects, similar to $S_{500\mu\rm m}$ limit given by \citet[][]{negrello_2010}{}{}.}

    \label{fig:S13mm}
\end{figure*}

 \citet[][]{negrello_2010}{}{}, found that sources that have 500\,$\mu$m fluxes densities above 100 mJy are expected to be gravitationally amplified (with close to 100 per cent accuracy). This can be further tested thanks to the large sample of red-{\it Herschel} sources and their 1-arcsec resolution ALMA observations considered in this work. As it can be seen in Fig. \ref{fig:S13mm} (see also Fig. \ref{fig:deb}), PLUM - Fields indeed dominate at $S_{\rm 500\,\mu m}\gtrsim100$\,mJy, with only two out of the eight sources in this regime cataloged as S - Fields.  However, deep higher angular resolution ($<1\,$arcsecond) data from the ALMA archive (e.g. Projects: 2016.1.00139.S P.I. R. Ivison and 2021.1.01628.S and 2022.1.00432.S, P.I. T. Bakx) show
evidence of strong lensing features in these two sources\footnote{These sources are HARPAS\_12 (R.A.: 00\,h 01\,m 24.8971\,s, Dec.: -35\,d 42\,m 12.046\,s), HARPAS\_2444
 (R.A.: 22\,h 42\,m 07.2143\,s, Dec.:-32\,d 41\,m 59.480\,s)}.
This, first, confirms the \citeauthor[][]{negrello_2010}{}{} criterion to identify lensed systems and, second, implies that we  cannot totally discard the possibility that some of the S - Fields are gravitationally amplified sources with no evident lensing features in the  $\sim1$\,arcsecond resolution ALMA images \citep[e.g.][]{oteo2017witnessing, Bakx_2024}.

\par We therefore explored the possibility of using the 1.3\,mm flux density as a criterion for the selection of gravitationally amplified galaxies. As it can be seen in Fig. \ref{fig:S13mm}, almost all sources with $S_{\rm 1.3mm}  \geq 13.0$\,mJy correspond to PLUM - Fields, with the exception of 4 S - Fields. Three of these S - Fields\footnote{HARPAS\_12 (R.A.: 00 h 01 m 24.8971 s, Dec.: -
35 d 42 m 12.046), HARPAS\_1789 (R.A.: 13\,h 33\,m 37.4892\, s, Dec.:24\,d 15\,m 39.326\,s) and HARPAS\_45 (R.A.: 00\,h 04\,m 55.4254\,s, Dec.:-33\,d 08\,m 12.653\,s)} are clearly gravitationally lensed at higher angular resolution ($<1$\,arcsecond) ALMA archival data (Projects 2016.1.00139.S, P.I. R. Ivison, 2017.1.00510.S, P.I. I. Oteo, 2021.1.01628.S, P.I. T. Bakx). The fourth source\footnote{HARPAS\_1861 (R.A.: 13\,h 39\,m 39.9034\,s, Dec.: 31\,d 22\,m 06.672\,s); NGP-203484 in \citet{Montaña_2021}. } has CO-line detections  \citep[][]{Montaña_2021}{}{} and show a factor $\sim 10$ departure from the $L_{\rm CO}$-FWHM relation, strongly suggesting it is also gravitationally amplified.

\par
\begin{figure*}
    \centering
    \includegraphics[width=0.9\textwidth]{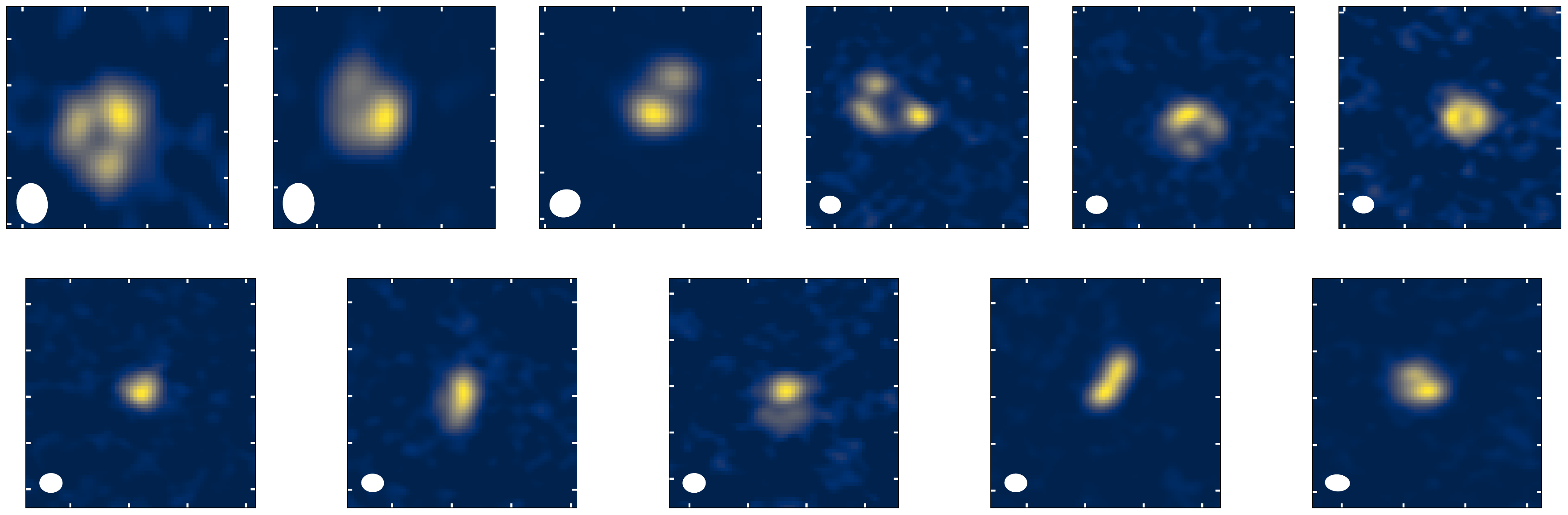}
    \caption{ 10 $\times$ 10\,square arcseconds postage stamps showing the ALMA 1.3\,mm observations of all the PLUM - Fields with $S_{\rm 1.3mm} > 13.0\, {\rm mJy}$. The white ellipse represents the synthesized beam (FWHM $\sim 1$\, arcsecond) for each observation. Most of these sources show evident gravitational lensing features. This potentially makes the $S_{\rm 1.3mm} > 13.0\, {\rm mJy}$ limit a selection criterion to identify gravitationally amplified DSFG. }
    \label{fig:lensed}
\end{figure*}

 Figure \ref{fig:lensed} shows all the PLUM - Fields with $S_{\rm 1.3mm} \geq 13.0$\,mJy, most of which (if not all) have evident lensing features.
Hence we postulate that DSFG with $S_{\rm 1.3\,mm}\geq13.0$\,mJy are most likely gravitationally amplified. This would also put a strong limit on the maximum luminosity of a dusty galaxy (and thus its SFR), as discussed in Section \ref{disc:LirandSFR}.

\subsection{Multiplicity}
\par Due to the low resolution of the \textit{Herschel} observations, it is important to understand if the properties estimated for the sources detected in these maps correspond to a single galaxy or are the result of blending multiple galaxies within the same beam. Higher angular resolution observations, e.g. with ALMA and more recently with \textit{JWST}, have confirmed cases of multiplicity in bright \textit{Herschel} sources. 
The $z = 6.34$ galaxy HFLS3 \citep[][]{Riechers_2013}{}{} is a clear example of such cases, which had long been considered an extreme SMG but \textit{JWST} has recently confirmed to be a high-$z$ interacting system of galaxies \citep[][]{Jones_2023}{}{}.

\par Considering the ALMA detections ($\geq 5\sigma$) within the \textit{Herschel} 500\,$\mu$m beam, we find that 475 out of 2,416 fields ($\sim 20$ per cent) are multiple systems with apparent separations $\geq 3$ arc-seconds between components\footnote{Recall that ALMA detections separated by $<3$ arc-seconds are considered single gravitationally lensed galaxies or potential closely interacting systems.}.  Of these 475 multiple systems, 87 per cent are doubles, 11 per cent are triples, and 2 per cent have four components. A fraction of these multiple systems may be physically associated galaxies (rather than line-of-sight projections), with some of the denser ones potentially corresponding to galaxy cluster progenitors (similar to those identified by \citealt{Oteo_2018} and \citealt{Jones_2023}).

\par Working with samples of similar galaxies (although significantly more limited in size), studies such as \citet{Ma_2019}, \citet{Greenslade_2020}, \citet{Montaña_2021}, \citet[][]{Bendo}, and \citet[][]{Cox_2024}{}{} reported that the multiplicity of the red-\textit{Herschel} sources ranges between 9 and 27 per cent, which is in good agreement with our results. \citet{Greenslade_2020} and \citet[][]{Montaña_2021}{}{} further suggested that, if their non-detections were multiple systems with individual components below the detection limit of the higher angular resolution observations, the multiplicity fraction might increase to $\sim 40 - 50$ per cent. If we assume that the red-\textit{Herschel} sources in our sample undetected in the ALMA 1.3\,mm maps are multiple systems, the multiplicity fraction would reach $\sim 37$ per cent.

\subsection{Redshift Distribution}\label{secc:disc_zdist}

Our results for the S - Fields show that these red-\textit{Herschel} galaxies have a median redshift of $2.78\pm0.03$, with 68 per cent of the population lying between $z = 2.13$ and 3.43, and a small tail that extends up to redshifts of $\sim6$ (Fig. \ref{fig:z}). This is a significantly different redshift space compared to that of the general population of \textit{Herschel}/SPIRE selected sources. For example, \citet[][]{Casey_2012}{}{} presented spectroscopic redshifts for 767 \textit{Herschel} galaxies (selected at 250, 350, and 500\,$\mu$m from HerMES) with a redshift distribution peaking at $z = 0.85$ and 731 sources at $z<2$. This highlights that the color selection of red sources within the general \textit{Herschel} population, indeed efficiently filters out the lower redshift ($z<2$) sources. 

Several works have presented photometric-redshift estimations of smaller samples of red-{\it Herschel} sources, for example: \citet{Ivison_2016} studied 109 SMG from the $H$-ATLAS with $850 \mu {\rm m}$ SCUBA-2/JCMT and $870 \mu {\rm m}$ LABOCA/APEX follow-up observations, and estimated a median redshift of 3.6; \cite{Ma_2019} compiled a sample of 300 {\it Herschel} $500\mu {\rm m}$-risers with 3.6 and 4.5 $\mu {\rm m}$ IRAC/{\it Spitzer} follow-up and found a median redshift of 3.7. Using a similar color selection to \citet{Ivison_2016}, \cite{Montaña_2021} presented 1.1 mm AzTEC/LMT observations of 100 $H$-ATLAS red galaxies, estimating a median redshift of 3.6. 

Compared to our results, these studies show redshift distributions systematically biased towards higher redshifts. This could be attributed to two main reasons: First, as mentioned above, our sample includes galaxies which are not as red as those selected by \citet{Ivison_2016}. Furthermore, all the sources in our sample are initially detected  in the 250$\mu$m band (SNR > $2.5\sigma$) and have flux densities above $4\sigma$ in at least one of the three SPIRE bands \citep[][]{Valiante_2016, Maddox_2018}{}{}, while the \citet{Ivison_2016} sample includes sources that are not necessarily detected in the shorter-wavelength bands, potentially selecting higher redshift galaxies (see {\sc bandflags} in \citealt{Ivison_2016}).

Second, the high sensitivity and angular resolution provided by the ALMA observations (compared to previous studies mostly based on single-dish telescope observations) allow us to better constraint the spectral energy distributions and photometric redshifts of these galaxies. Additionally, the different methods used to estimate photometric redshifts might introduce systematic discrepancies between the works. However, as shown in Fig. \ref{fig:compzspec_phot} our photometric redshifts show a good agreement with the available spectroscopic redshifts for the sample.

\par Beyond the studies of red-\textit{Herschel} galaxies, other works have focused on the properties of high-$z$ DSFG selected at longer wavelengths. For example, \citet{Simpson_2020} presented ALMA follow-up observations of $\sim 180$ SCUBA-2 sources with $S_\mathrm{850 \mu\mathrm{m}}>6.2$\,mJy (i.e. of order $S_{500\mu\mathrm{m}} = 20-50$\,mJy, the $500\mu\mathrm{m}$ detection limit of our sample), and estimated a median redshift of 2.87$\pm$0.08 for their sample. Although dependent on the specific detection limit and selection wavelength ($850 - 1300\mu{\rm m}$), different studies of the DSFG population with ALMA have measured redshift distributions with median values between 2.5 and 3, mainly estimated using Optical-FIR photometric redshifts and a few with spectroscopic confirmations \citep[e.g.][]{HATSUKADE_2018,Franco_2018,Dudze_2020}. 

The general agreement between these studies and our estimations suggests that the selection of red-{\it Herschel} sources overlaps with that of the classic SMG population selected at longer wavelengths, although biased towards higher luminosities (as seen in Fig. \ref{fig:Lirz}) due to the shallower depth of the {\it Herschel} surveys and higher sensitivity to the rarest brightest and highest-redshift systems given the larger areas covered by the {\it Herschel} blank-field surveys.

\par For the sample of PLUM - Fields, which are likely dominated by strongly lensed systems (Fig. \ref{fig:lensed}), we estimate a median redshift of $3.28\pm0.12$, with 83 (i.e. $\sim60$ per cent) at $z>3$. This is slightly higher that the median redshift found for the S-Fields and, as it was previously mentioned, this could be the result of a bias introduced by gravitational lensing effects. 
Compared to \cite{urquhart+2022}, who presented spectroscopic redshifts ($1.41 < z_{\rm spec} < 4.53$) for 71 gravitationally lensed {\it Herschel}-galaxies, our sample shows a higher median redshift (3.28 vs 2.75). This is expected since their selection only included sources with $S_{\rm 500 \mu m} >80\,$mJy and $z_{\rm phot}>2$, without any further color restrictions. On the other hand, \citet{Reuter_2020} reported a spectroscopically-derived median redshift of $3.9\pm0.2$ for the 81 strongly gravitationally lensed DSFG in the SPT-SZ Survey, selected at 1.4\,mm. This larger value, compared to our estimate for PLUM - Fields, can be attributed to the longer wavelength observations used to select their sample, which are expected to select, on average, higher redshift sources \citep[e.g][]{Zavala_2014,Bethermin_2015}{}{}.

\subsection{Infrared Luminosities and Star Formation Rates\label{disc:LirandSFR}}
\par The infrared luminosities of the S-Fields have a median of ($8.9\pm 0.2)\times 10^{12}$\,L$_\odot$. These luminosities are consistent with what is expected for this population and lead to star formation rates of $\sim$1,300\,M$_\odot$\,yr$^{-1}$. Several studies of red-{\it Herschel} galaxies in the \textit{H}-ATLAS field find median infrared luminosities between $9\times10^{12}$\,L$_\odot$ and $2\times10^{13}$\,L$_\odot$ \citep[][]{Ivison_2016,Ma_2019,Montaña_2021} with SFR of $\sim$ 800 -- 2,100\,M$_\odot$\,yr$^{-1}$. Similarly, \cite{Greenslade_2020} studied a sample of 34 red-{\it Herschel} sources from HerMES and found infrared luminosities of $\sim (2 - 6) \times10^{13}$\,L$_\odot$. The higher luminosities in these studies might reflect samples biased towards brighter sources, particularly in the case of \cite{Greenslade_2020}, with a selection criterion of $S_{500\mathrm{\,\mu m}}>60$\,mJy, compared to $>35$\,mJy in \citet[][]{Ivison_2016,Ma_2019,Montaña_2021}, and $>20$\,mJy in this work (after deblending). Moreover, the studies based on single-dish telescope observations, might be more contaminated by lensed sources than our S-Field sample, that benefits from high angular resolution ALMA data.

\par
\begin{figure}
    \centering
    \includegraphics[width=\columnwidth]{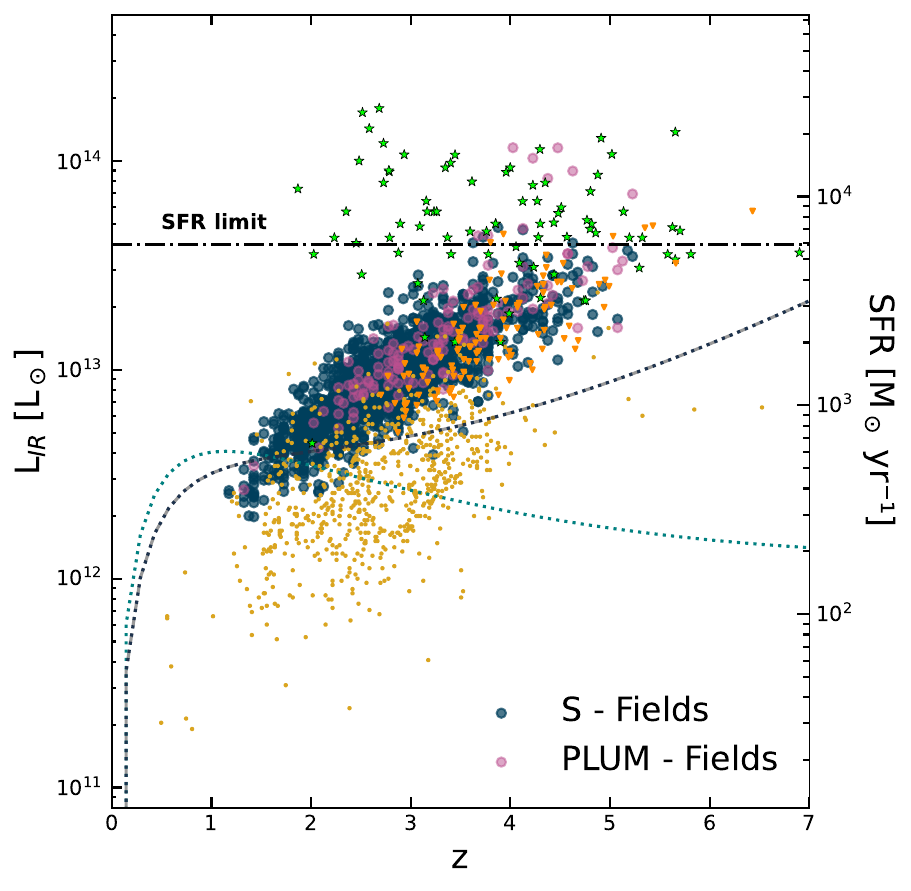}
    \caption{Infrared luminosities (left $y$-axis) and SFR (right $y$-axis) of our samples as a function of redshift. PLUM - Fields, being gravitationally amplified or possible close interacting galaxies, reach larger values of $L_{\mathrm{IR}}$ at higher redshifts. Yellow dots correspond to $\sim 700$ SCUBA-2 SMG with ALMA follow-up observations \citep{dudzevivciute2020}. The green stars are ALMA follow-up of 1.4 mm SPT selected sources \citep[not corrected for gravitatinal lensing amplification,][]{Reuter_2020}. To convert $L_{\mathrm{FIR}}$ from \citet[][]{Reuter_2020}{}{} to $L_{\mathrm{IR}}$, we use $L_{\mathrm{IR}}\simeq1.3\times L_{\mathrm{FIR}}$ \citep{Miettinen}. Orange triangles correspond to the SCUBA-2 follow-up of ``ultra-red'' {\it H}-ATLAS sources from \citet[][]{Ivison_2016}{}{}.  Also shown are the redshift tracks of two modified black bodies with emissivity index $\ \beta=2$, dust temperature $T_{\mathrm{dust}}=45$\,K and: $S_{500\mathrm{\mu m}}=20$\,mJy (i.e. lower flux density after deblending, black dotted curve), and $S_{1.3\mathrm{\,m m}}=1$\,mJy (i.e. our average ALMA 5$\sigma_{\mathrm{1.3\,mm}}$ detection limit, light blue dotted curve). 
    From this figure, an upper limit of $\sim 6,000$\,M$_\odot$\,yr$^{-1}$ (horizontal dot-dashed line) can be established to the SFR of the S-Fields. The nine PLUM - Fields that lie above this limit are shown in Fig. \ref{fig:lensed}. }
    \label{fig:Lirz}
\end{figure}

 Figure \ref{fig:Lirz} shows the $L_{\rm IR}$ (and SFR) of our sources as a function of redshift, and reveals something interesting about the maximum SFR of the sample. We note that all sources with $L_{\rm IR} \gtrsim 4\times10^{13}$\,L$_\odot$ correspond to PLUM-Fields or can be identified as gravitationally lensed galaxies using high angular resolution ($<1$\, arcsecond) ALMA archival data (See Section \ref{sec:fluxandcolors}). This suggests that a single galaxy cannot reach this luminosity levels unless it is gravitationally amplified. In other words, there seems to exist a redshift independent SFR upper limit for SMG around (or below) $\sim6,000$\,M$_\odot$\,yr$^{-1}$.  This is consistent with results from other studies, for example, \citet[][]{Barger_2014}{}{}, using a sample of 850\,$\mu$m-selected SMG in an area of 400\,square\,arcminutes, found a peak at $\sim2,000$\,M$_\odot$\,yr$^{-1}$ in the SFR distribution with values up to $\sim6,000$\,M$_\odot$\,yr$^{-1}$. Galaxies with reported SFR above this value (e.g. \citealt{Rowan-Robinson_2018}) are, therefore, most likely gravitationally amplified or multiple galaxies blended within a telescope beam. However, this limit only represents a loose constraint (probably being overestimated) since some of the S - Fields might still be contaminated by gravitational lensing. 

\par Finally, we note that 87 S-Fields have SFR $\gtrsim3,000$\,M$_\odot$\,yr$^{-1}$. These systems might represent some of the most extreme (unlensed) galaxies found to-date. However, although our sample was cleaned from AGN contamination (Section \ref{sec:samplecleaning}), we cannot rule out the possibility that some of these extreme SFR may be overestimated ($\sim 10-30$\,per\,cent, \citealt[][]{Casey2014}{}{}) by the presence of a dust obscured AGN. In fact, \citet[][]{Rowan-Robinson_2018}{}{} found evidence of an AGN in $\sim50$ per cent of the 38 \textit{Herschel} extreme starbursts (SFR$>5,000\mathrm{\,M_{\odot}\,yr^{-1}}$) in their sample. In addition some of these extreme galaxies could still be gravitationally lensed or multiple systems not resolved by the 1\,arcsecond ALMA resolution, as mentioned above.

\subsection{Gas and Dust Mass}
\par As described in Section \ref{sec:gas_masses}, we estimate gas masses we following the method presented in \citet[][]{Scoville_2016}{}{}, which considers optically thin dust continuum emission as a tracer of the interstellar medium mass (see Section \ref{sec:gas_masses}). The medians for our samples are $M_{\mathrm{gas}}=(4.1\pm0.1)\times 10^{11}$\,M$_{\odot}$ for the S - Fields and $M_{\mathrm{gas}}=(5.7 \pm 0.3)\times 10^{11}$\,M$_{\odot}$ for the PLUM - Fields. The galaxies in our sample are therefore massive systems, with large enough gas reservoirs to sustain their high SFR ($\sim 1000\, {\rm M}_\odot {\rm yr^{-1}}$) for tens and even hundreds of millions years, as discussed below.

\par Different authors have used direct measurements of CO (1-0), or extrapolations from CO mid-J transitions, to study the molecular gas content in samples of a few to tens of infrared luminous DSFG \citep[e.g.][]{bothwell,Aravena_2016,yang_2017,Birkin_2020,Ikarashi_2022,Berta_2023}{}{}, estimating gas masses of $M_{\rm gas}\sim10^{10-11}$\,M$_{\odot}$.
These values are in broad agreement with the estimates presented in this work, particularly when considering the different methodologies, assumptions on the $\alpha_{\mathrm{CO}}$ (typically from 0.8 to 6.5), and different CO line ratios, which could introduce large uncertainties. For example, adopting $\alpha_{\mathrm{CO}}=1 \mathrm{M_\odot(K \,km\ ,s^{-1}\,pc^{2})^{-1}}$ (a commonly used value for SMG, e.g. \citealt{bothwell, Birkin_2020}) our gas mass determinations decrease to $M_{\mathrm{gas}}=(6.3\pm0.2)\times 10^{10}$\,M$_{\odot}$ for the S - Fields and $M_{\mathrm{gas}}=(8.8 \pm 0.5)\times 10^{10}$\,M$_{\odot}$ for the PLUM - Fields.

\par Finally, our dust mass estimations are consistent with massive DSFG at high redshift \citep[e.g][]{Casey_2019,Ma_2019}{}{} with median values of $\sim10^{9}$\,M$_{\odot}$. 
This result confirms the large dust content of these galaxies that obscures most of the star formation activity in this high-$z$ population.

\subsection{Depletion times and Star formation efficiencies}
\par
\begin{figure}
    \centering
    \includegraphics[width=\columnwidth]{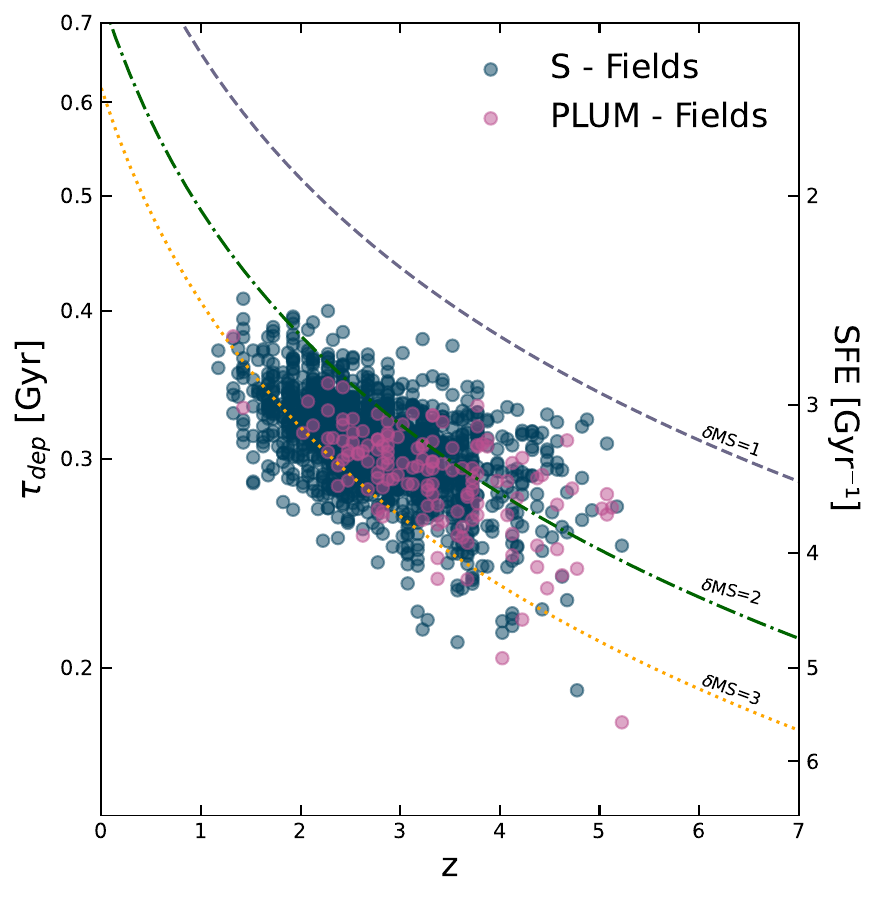}
    \caption{Depletion time ($\tau_{\mathrm{dep}}$) as a function of redshift for S - Fields and PLUM - Fields samples. The curves depict the $\tau_{\mathrm{dep}}-z$ relationship \citep[][]{tacconi_2018} for galaxies in the main sequence ($\delta \mathrm{ MS}=1$, purple dashed curve), and starburst type galaxies $\delta \mathrm{ MS}=2$ \citep[][green dash-dot curve]{Elbaz_2011}{}{} and $\delta \mathrm{ MS}=3$ \citep[][yellow dotted curve]{Franco_2020}{}{}. 
    The bulk of our samples has depletion times consistent with a starburst population.}
    \label{fig:tdep}
\end{figure}

  The depletion times ($\tau_{\mathrm{dep}} = M_{\mathrm{gas}}/\mathrm{SFR}$), which are independent of gravitational amplification to a first order, can give clues about the processes behind the extreme SFR in these galaxies. We use the values estimated in Sections \ref{subsec:red_lum_sfr} and \ref{sec:gas_masses} to determine $\tau_{\mathrm{dep}}$ (and star formation efficiencies (SFE), i.e. $1/\tau_{\mathrm{dep}}$) for the sources in our sample. Figure \ref{fig:tdep} shows our depletion time estimates versus redshift, along with the expected evolution for main-sequence star-forming galaxies and starburst type galaxies \citep[][]{tacconi_2018}{}{}. Our $\tau_{\mathrm{dep}}$ are systematically shorter than those expected for main-sequence galaxies and consistent with the starburst population. These results, however, should be taken with caution since the $M_{\mathrm{gas}}$ are inferred from the ALMA 1.3\,mm photometry and are not totally independent from the SED-based SFR determinations.

\par The inferred depletion times for the S - Fields cover a range from 0.2\,Gyr to 0.4\,Gyr. These values are also consistent with those found in the literature for similar samples, where the depletion times are between 0.1\,Gyr and 1\,Gyr \citep[][]{Hatsukade_2015, Pantoni_2021,Berta_2023,Castillo_2023}{}{}, but lower compared to those found for nearby galaxies \citep[$z<1$ and $\tau_{\mathrm{dep}}> 1$ Gyr, e.g][]{Bigiel_2008, Leroy_2008,genzel_2010}{}{}. If we scale our gas mass determinations to assume $\alpha_{\mathrm{CO}} = 1\,\mathrm{M_\odot (K\,km\,s^{-1}\,pc^{2})^{-1}}$, our depletion times are reduced even further (30 to 60\,Myr). These values are in agreement with those reported by the extreme SMG population reported to have high star formation efficiency \citep[][]{Aravena_2016,yang_2017,Ikarashi_2022}{}{}.

\par Considering that $\sim73$\,per\,cent of the sources in our sample are individual (at 1\,arcsecond resolution), we suggest that the processes that increase the SFE in these galaxies act at scales of $\lesssim8.5\,$kpc as are near-coalescence mergers and other physical processes such as gravitational instabilities and gas turbulence.

\section{Summary and Conclusions}\label{sec:summary}
Using ALMA archival data  at 1.3\,mm with 1\,arcsecond angular resolution, we studied a sample of $\sim$ 3,000 red-\textit{Herschel} sources selected in the SPIRE/\textit{Herschel} bands ($S_{\mathrm{250 \mu m}}<S_{\mathrm{350 \mu m}}<S_{\mathrm{500 \mu m}}$). We found 2,416 ALMA maps with detections ($\geq5\sigma$), of which 1,762 show a single point-like source (S - Fields; 73\,per\,cent), 475 fields with multiple sources (M - Fields; 20\,per\,cent), 137 classified as potential gravitationally lensed or closely interacting systems (PLUM - Fields; 5\,per\,cent), and 43 identified as low-$z$ interlopers or AGN. All sources with $S_{1.3\rm mm} \geq 13.0\,$mJy show features of strong gravitational amplification. This implies a redshift-independent limit of $\sim6,000$\,$\mathrm{M_{\odot}\,yr^{-1} }$ to the maximum SFR of SMG and provides a simple photometric criterion to identify lensed systems in future millimeter surveys.

\par Combining the SPIRE/\textit{Herschel} photometry with the ALMA band 6 imaging, we conducted an SED fitting analysis and infer the physical properties for the S - Field and PLUM - Field samples. We found a  median redshift of 2.78$\pm$0.03 for the S - Fields and 3.28$\pm$0.12 for the PLUM - Fields. The slightly higher redshift of the PLUM-fields could be attributed to a selection bias, since 
the probability of a source being gravitationally lensed increases as a function of its redshift. 
When comparing the redshift distribution of the single sources in our sample with earlier studies of red-{\it Herschel} sources, we found that our distribution is shifted towards smaller redshifts. This may be due to the fact that our sample is not as ``red'' as those presented in previous works which, given the smaller size of their samples, targeted the most extreme sources in terms of color and brightness.
The redshift distribution of the red-{\it Herschel} sources confirmed to be single systems (i.e. S - Fields) is similar to those estimated for galaxies selected at 850\,$\mu$m or 1.1\,mm from single-dish surveys. The larger area of the {\it Herschel} surveys, however, increases the probability of identifying the rarest, the brightest, and the highest-redshift systems.

\par The median IR luminosities for the S - Fields and PLUM - Fields of $8.9\times 10^{12}$ and $1.3\times 10 ^{13}$\,L$_{\odot}$, correspond to SFR of $\sim1,300$ and $\sim2,000$\,$\mathrm{M_{\odot}\,yr^{-1}}$, respectively. This is in line with previous studies and confirms the extreme nature of these SMG, which also show high gas and dust content with $M_{\rm gas}\sim4.9\times 10^{11}\,\mathrm{M_\odot}$ and $M_{\rm dust}\sim2.6\times 10^{9}\,\mathrm{M_\odot}$. Combining these measurements, we found average depletion times of $\tau_{\mathrm{dep}}=0.30\pm0.03$\,Gyr, consistent with what has been measured for other SMG, implying high SFE compared to the main-sequence population. We stress, however, that these results should be taken with caution until a more direct measurement of the gas masses become available.

\par 
Since most of our sources ($\sim73$\,per\,cent) are individual systems at the current 1 arcsecond resolution, we suggest that the processes increasing the SFE in SMGs act at scales of $\lesssim8.5\,$kpc. This rejects early-mergers with separations of several tens of kiloparsecs as the main drivers of the extreme SFR of these galaxies, leaving the door open to mergers near coalescence and other physical processes such as gravitational instabilities and gas turbulence.

These results shed light on the nature of the red-\textit{Herschel} galaxies, which are tracing the most extreme starbursts from the end of the epoch of reionization to cosmic noon. Finally, the legacy catalogs from this work provide excellent targets for follow-up observations aimed at understanding the resolved properties of the most extreme star-forming galaxies and at finding high redshift overdensities of massive dusty galaxies.

\section*{Acknowledgements}
We thank to the referee for insightful comments that have improved
the manuscript. We also want to thank Olga Vega, Yalia Divakara Mayya, Manuel Zamora-Avilés, Elena Terlevich and Eric F. Jiménez-Andrade for their helpful discussions and comments, and to Micha\l{} Micha\l{}owski for insightful suggestions on the flux-deblending process. MQR would like to thank the Consejo Nacional de Humanidades Ciencias y Technolog\'ias (CONAHCYT) for her PhD grant. This work has been supported by the CONAHCYT through projects A1-S-45680 and CB 2016 - 281948. JAZ acknowledge funding from JSPS KAKENHI grant number KG23K13150. 
The \textit{Herschel}-ATLAS is a project from \textit{Herschel}, which is an ESA space observatory with science instruments provided by European-led Principal Investigator consortia and with important participation from NASA.
This paper use the following ALMA projects: ADS/JAO.ALMA\#2016.1.00087.S,  ADS/JAO.ALMA\#2016.1.00139.S, ADS/JAO.ALMA\#2017.1.00510.S, ADS/JAO.ALMA\#2018.1.00489.S, ADS/JAO.ALMA\#2018.1.00526.S, ADS/JAO.ALMA\#2021.1.01628.S and ADSJAO.ALMA\#2022.1.00432.S. ALMA is a partnership of ESO (representing its member states), NSF (USA), and NINS (Japan), together with NRC (Canada), MOST and ASIAA (Taiwan), and KASI (Republic of Korea), in cooperation with the Republic of Chile. The Joint ALMA Observatory is operated
by ESO, AUI/NRAO, and NAOJ. Funding for the Sloan Digital Sky Survey V has been provided by the Alfred P. Sloan Foundation, the Heising-Simons Foundation, the National Science Foundation, and the Participating Institutions. SDSS acknowledges support and resources from the Center for High-Performance Computing at the University of Utah. SDSS telescopes are located at Apache Point Observatory, funded by the Astrophysical Research Consortium and operated by New Mexico State University, and at Las Campanas Observatory, operated by the Carnegie Institution for Science. The SDSS web site is \url{www.sdss.org}.
SDSS is managed by the Astrophysical Research Consortium for the Participating Institutions of the SDSS Collaboration, including Caltech, the Carnegie Institution for Science, Chilean National Time Allocation Committee (CNTAC) ratified researchers, The Flatiron Institute, the Gotham Participation Group, Harvard University, Heidelberg University, The Johns Hopkins University, L’Ecole polytechnique fédérale de Lausanne (EPFL), Leibniz-Institut für Astrophysik Potsdam (AIP), Max-Planck-Institut für Astronomie (MPIA Heidelberg), Max-Planck-Institut für Extraterrestrische Physik (MPE), Nanjing University, National Astronomical Observatories of China (NAOC), New Mexico State University, The Ohio State University, Pennsylvania State University, Smithsonian Astrophysical Observatory, Space Telescope Science Institute (STScI), the Stellar Astrophysics Participation Group, Universidad Nacional Autónoma de México, University of Arizona, University of Colorado Boulder, University of Illinois at Urbana-Champaign, University of Toronto, University of Utah, University of Virginia, Yale University, and Yunnan University.
\section*{Data Availability}
Catalogues are available in supplementary data at MNRAS online. The \textit{Herschel} Astrophysical Terahertz Large Area Survey (\textit{H}-ATLAS) catalogs and maps are available in: \url{https://www.h-atlas.org/public-data}. The ALMA data band 6 is available in 
\url{ https://data.nrao.edu/portal/} for the MS files and in \url{https://almascience.eso.org/aq/} for the maps.



\bibliographystyle{mnras}
\bibliography{mnras_template_ARXIV_v1} 




\appendix

\section{Sources classified as Active Galactic Nuclei and low-redshift galaxies\label{sec:apex1}}
\par Table \ref{tab:agn_lowz} lists the 43 fields removed from our sample due to their association with Active Galactic Nuclei (86 per cent) and low-$z$ galaxies (14 per cent) using the MILLIQUAS and the $17^{\rm th}$ Data Release of the Sloan Digital Sky Survey. These fields represent only 2\,per\,cent of the total sample of red {\it H}-ATLAS sources identified in the public ALMA archive.

\begin{table*}
\centering
\begin{tabular}{ccccccccc}
 \hline
IAU Name&Type & Catalog & Catalogs Names & R.A. & Dec. & $z$ \\
\hline
HATLASJ000106.6-320638&AGN & MILLIQUAS &2QZ J000106.9-320642& 0.277693 & -32.110526 &  & \\
 HATLASJ005802.3-323420&AGN & MILLIQUAS &PKS 0055-328   & 14.509378 & -32.572155 &  & \\
HATLASJ014310.0-320056 &AGN & MILLIQUAS & PKS 0140-322  & 25.791662 & -32.015671 & 0.372 \\      
HATLASJ014503.4-273333
 &AGN & MILLIQUAS & PKS 0142-278 & 26.263985 & -27.559147 &  & \\
HATLASJ084924.5+014712&AGN & MILLIQUAS & SDSS J084924.52+014713.1 & 132.352127 & 1.786842 &  & \\
HATLASJ090910.1+012134&AGN & MILLIQUAS & PKS 0906+01 & 137.292358 & 1.359685 &  & \\
HATLASJ090940.2+020004& AGN* & NED &PKS 0907+022 & 137.417755&2.001126 & 1.601 & \\
HATLASJ113320.2+004053& AGN & MILLIQUAS & PKS 1130+009 & 173.334182 & 0.681598 &  & \\
HATLASJ092654.4-002149& AGN & MILLIQUAS & PKS 1148-00 & 177.681732 & -0.398445 &  & \\
HATLASJ120741.7-010636& AGN & MILLIQUAS & PKS 1205-008 & 181.923797 & -1.110082 &  & \\
HATLASJ121758.7-002946& AGN & MILLIQUAS & PKS 1215-002 & 184.494980 & -0.496351 & 0.419 & \\
HATLASJ125346.9+232733 & AGN & MILLIQUAS & SDSS J125347.08+232726.8 & 193.445602 & 23.459068 &  & \\
HATLASJ125516.4+281827& AGN & MILLIQUAS & SDSS J125516.40+281828.0 & 193.818436 & 28.307546 &  & \\
HATLASJ125728.1+292711 & AGN & MILLIQUAS & SDSS J125728.05+292715.1 & 194.367233 & 29.453180 &  & \\
HATLASJ125757.3+322930& AGN & MILLIQUAS & B2 1215+32 & 194.488815 & 32.491760 & 0.806 &\\
HATLASJ130129.0+333703 & AGN & MILLIQUAS & 5C 12.170 & 195.370910 & 33.617420 &  & \\
HATLASJ131028.7+322044 & AGN & MILLIQUAS & B2 1308+32 & 197.619659 & 32.345508 & 0.998 \\
HATLASJ131059.2+323331& AGN & MILLIQUAS & MG J1310+3233 & 197.746750 & 32.558670 &  & \\
HATLASJ131736.4+342518 & AGN & MILLIQUAS & B2 1315+34A & 199.401718 & 34.421684 &  & \\
HATLASJ131947.3+292640& AGN & MILLIQUAS & SDSS J131946.93+292640.4 & 199.947204 & 29.444395 &  & \\
HATLASJ132952.9+315410& AGN & MILLIQUAS & 87GB 13275+3209 & 202.4702604 & 31.902751 &  & \\
HATLASJ133038.1+250901 & AGN & MILLIQUAS & 3C 287.0 & 202.658569 & 25.150234 &  & \\
HATLASJ133108.4+303034& AGN & MILLIQUAS & 3C 286.0 & 202.785171 & 30.509453 & 0.85 \\
HATLASJ133120.4+291200& AGN & MILLIQUAS & SDSS J133120.40+291200.6 & 202.835144 & 29.199993 &  & \\
HATLASJ133307.4+272518 & AGN & MILLIQUAS & CGRaBS J1333+2725 & 203.280807 & 27.421665 & 0.731 \\
HATLASJ134131.1+235043 & AGN & MILLIQUAS & SDSS J134131.14+235043.3 & 205.379699 & 23.845237 &  & \\
HATLASJ134208.4+270933& AGN & MILLIQUAS & RRS IV 4 & 205.535141 & 27.159134 &  & \\
HATLASJ140729.5+011216 & AGN & MILLIQUAS & SDSS J140729.39+011218.5 & 211.873001 & 1.204640 &  & \\
HATLASJ141004.7+020306
 & AGN & MILLIQUAS & PKS 1407+022 & 212.519637& 2.0518739 &  & \\
HATLASJ142121.3-001136& AGN & MILLIQUAS & SDSS J142121.54-001138.9 & 215.339096 & -0.193460   & \\
HATLASJ144424.5-004454
& AGN & MILLIQUAS & SDSS J144424.04-004454.5 & 221.102218 & -0.748408 &  & \\
HATLASJ145146.1+010608& AGN & MILLIQUAS & MASIV J1451+0106 & 222.942489 & 1.102491 &  & \\
HATLASJ222321.6-313701 & AGN & MILLIQUAS &J222321.63-313702.1  & 335.839965 & -31.616873 &  & \\
HATLASJ224838.6-323551 & AGN & MILLIQUAS & PKS 2245-328      & 342.160827 & -32.597446 &  & \\
HATLASJ231859.8-294738 & AGN & MILLIQUAS & 2QZ J231859.9-294736 & 349.749207 & -29.793848 &  & \\
HATLASJ235935.3-313343& AGN & MILLIQUAS & PKS 2357-318 & 359.897247 & -31.562063 & 0.991 & \\
HATLASJ235347.4-303746&AGN&MILLIQUAS&PKS 2351-309&358.447571&-30.629381\\


\hline
HATLASJ131911.4+312732 & Low-$z$ galaxy & SDSS & J131911.29+312725.4  & 199.797500 & 31.458898 & 0.256 \\
HATLASJ132545.8+264126& Star* &NED &  J132545.54+264123.6 & 201.440674 & 26.690624 &  & \\
HATLASJ134339.0+352335& Low-$z$ galaxy & SDSS & J134338.97+352337.3 & 205.91231 & 35.392933 & 0.2 \\
HATLASJ134857.9+245357 & Low-$z$ galaxy &  SDSS&  J134857.90+245356.4 & 207.24115 & 24.899031 & 0.6 &\\
HATLASJ142111.9-013845& Low-$z$ galaxy &SDSS  &  J142111.98-013843.9 & 215.299987 & -1.646102 & 0.279 &\\
HATLASJ234758.4-331918 & Low-$z$ galaxy&SDSS  & TGS492Z213 & 356.993225 & -33.321781 & 0.21   \\       
\hline
    \end{tabular}

\begin{minipage}{15cm}
\small *Classified by NASA/IPAC Extragalactic Database.
\end{minipage}
\caption{Fields from our sample that match (at $\leq$ 2 arcsecond) a source in the The Million Quasars (MILLIQUAS) Catalog \citep[]{Milliquas} and the $17^{\rm th}$ Data Release of the Sloan Digital Sky Survey \citep[SDSS,][]{SDSS_DR17}. 
 Columns 5 and 6 are the coordinates of ALMA maps centers. }
    \label{tab:agn_lowz}
\end{table*}


\bsp	
\label{lastpage}
\end{document}